\documentclass{elsart}

 \usepackage{graphicx}

\usepackage{amssymb}
\usepackage{natbib}

\begin{document}

\begin{frontmatter}
\title{Anyons and the quantum Hall effect - a pedagogical review}
\author{Ady Stern}
\address{Department of Condensed Matter Physics, Weizmann Institute of Science, Rehovot 76100, Israel}
\ead{Adiel.Stern@Weizmann.ac.il}






\begin{abstract}

The dichotomy between fermions and bosons is at the root of many physical phenomena, from metallic conduction of electricity to super-fluidity, and from the periodic table to coherent propagation of light. The dichotomy originates from the symmetry of the quantum mechanical wave function to the interchange of two identical particles. In systems that are confined to two spatial dimensions particles that are neither fermions nor bosons, coined "anyons", may exist. The fractional quantum Hall effect offers an experimental system where this possibility is realized. In this paper we present the concept of anyons, we explain why the observation of the fractional quantum Hall effect almost forces the notion of anyons upon us, and we review several possible ways for a direct observation of the physics of anyons. Furthermore, we devote a large part of the paper to non-abelian anyons, motivating their existence from the point of view of trial wave functions, giving a simple exposition of their relation to conformal field theories, and reviewing several proposals for their direct observation.
\end{abstract}

\end{frontmatter}

There are two basic principles on which the entire formidable world of non-relativistic quantum mechanics resides. First, the world is described in terms of wave functions that satisfy Schroedinger's wave equation. And second, these wave functions should satisfy certain symmetry properties with respect to the exchange of identical particles. For fermions the wave function should be anti-symmetric, for bosons it should be symmetric. It is impossible to overrate the importance of these symmetries in determining the properties of quantum systems made of many identical particles. Bosons form superfluids, fermions form Fermi liquids. The former may carry current without dissipating energy, the latter dissipate energy. The periodic table of elements, and with it chemistry and biology, looks the way it does because electrons are fermions. And radiation may propagate in a coherent way since photons are bosons. 

Given this set of reminders, it should be clear that finding particles that are neither fermions nor bosons is an exciting development. Finding them as a theoretical construct is exciting enough, as realized by Leinaas and Myrheim \cite{Leinaas77} and by Wilczek \cite{Wilczek82a}. Having them in the laboratory, open for experimental investigation with an Amperemeter and a Voltmeter is plain wonderful. Luckily, two dimensional electronic systems subjected to a strong magnetic field provide both the theoretical and experimental playground for such an investigation, all through the Quantum Hall Effect \cite{Prange90,DasSarma97,Klitzing80,Tsui82}. 

This paper is aimed at reviewing the physics of Anyons, particles whose statistics is neither fermionic not bosonic, and the way it is manifested in the quantum Hall effect. We will start with introducing the basic characters of this play - the Quantum Hall effect, the Aharonov-Bohm effect \cite{ABeffect59} and (more briefly) the Berry phase\cite{Berry84}. We will then show why the mere experimental observation of the quantum Hall effect, coupled with very basic principles of physics, forces us to accept the existence of excitations that effectively are Anyons, and how it raises the distinction between abelian and non-abelian anyons. Following that, we will explore what the experimental consequences of anyonic quantum statistics may be, with a strong emphasis on interference phenomena. After this subject is covered, we will raise one level in complexity, and introduce non-abelian anyons\cite{MR}, looking first at the general concept, then at the simplest example, that of the $\nu=5/2$ quantum Hall state, and finally at the more complicated example of the Read-Rezayi states\cite{Read99}. 

Some subjects will be left out. In particular, the relation of anyons to topological quantum computation \cite{Kitaev97}, to topological quantum field theories and to group theory are all covered at great length in a recent review article whose list of authors has some overlap with the corresponding list of the present paper\cite{RMP}. These subjects will not be repeated here. Left out are also realizations of anyons in systems out of the realm of the quantum Hall effect, the relation of anyons to exclusion statistics (a generalized Pauli principle for anyons\cite{PhysRevLett.67.937,mathieu2,Ardonne01,Bouwknegt99,PhysRevLett.81.1929}) and studies of statistical physics of anyons in abstract models. 

The paper attempts to be pedagogical, and should be accessible to graduate students, both theorists and experimentalists.

\section{The quantum Hall effect}
Let us think of electrons on an infinite conducting strip of width
$w$ on the $x-y$ plane, subjected to a perpendicular magnetic field $B{\hat z}$. If  an electronic current is flowing in the $\hat
x$-direction, and the strip is confined in the $\hat y$-direction,
a Hall voltage $V_H$ develops in the $\hat y$-direction.
Classically, this voltage is calculated by a seemingly bullet-proof argument: for the current to flow in a
straight line, the Lorenz force originating from the magnetic
field should be cancelled by the force originating from the
gradient of the Hall voltage. Thus, one expects,
\begin{equation}
\frac{e}{c}vB=\frac{V_H}{w} \label{classicalHall1}
\end{equation}
where $v$ is the electron's velocity, $e$ is the electron charge and $c$ is the speed of light. Since the current is
$I=nevw$, with $n$ being the electron's density, we get
\begin{equation}
\frac{V_H}{I}=\frac{B}{nec}\label{classicalHall2}
\end{equation}
This ratio of the Hall voltage to the current is known as the Hall
resistance, denoted by $R_{yx}=-R_{xy}$ or $R_H$.  The simple-minded line of thoughts goes even further: once the force that results
from the Hall voltage cancels the force that results from the
magnetic field, there will be no
other effect of the magnetic field. Thus, the longitudinal
voltage, the voltage drop parallel to the current, will be
independent of the magnetic field. The ratio of this voltage to
the current is the longitudinal resistance $R_{xx}$.

Classical arguments are not the whole story. There are two other
things to add: real-life experiments and quantum mechanics. Both
contradict the classical arguments in a very fundamental way.
Before we get to  the
quantum Hall effect, we should pay homage to an even earlier contradiction - the sign of the Hall
voltage. Indeed, while classical physics relates this sign to the
direction of the Lorentz force that acts on the negatively charged
electrons, experiments show a different sign in different
materials. Quantum mechanics, in one of its greatest triumphs
in solid state physics, explains this varying sign in terms of
the theory of bands, and the notion of holes.

\begin{figure}
\includegraphics[width= 0.9\linewidth]{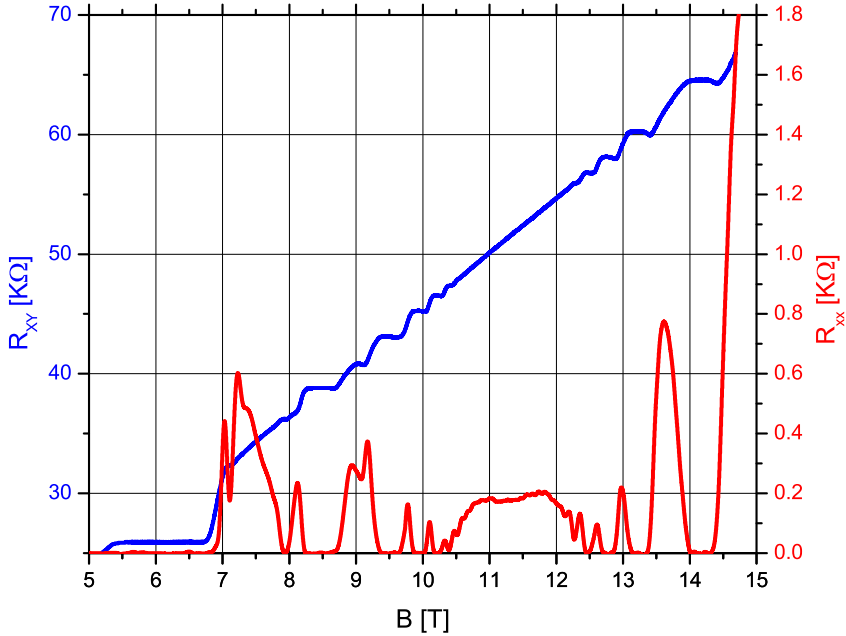}
\caption{\label{QHE} The quantum Hall effect. When the Hall resistance is measured as a function of magnetic field plateaus at quantized values are observed. In regions of the magnetic field where the Hall resistance is in a plateau, the longitudinal resistance vanishes (sample grown by Vladimir Umansky and Measured by Merav Dolev, the Weizmann Institute of Science). }
\end{figure}

Several decades after band theory introduced the holes,
the quantum Hall effect was discovered\cite{Prange90,DasSarma97,Klitzing80,Tsui82}. Look at Fig. (\ref{QHE}): when the Hall effect is
measured in high mobility two dimensional electronic systems at
low temperatures, the Hall resistance is not linear in magnetic
field, in contrast to what  Eq. (\ref{classicalHall2}) suggests. Rather, it shows
steps. The magnetic field at which a step starts or ends varies
between samples, but the value of $R_H$ at the steps is universal.
At the steps,
\begin{equation}
R_H=\frac{h}{e^2}\frac{1}{\nu} \label{quantumrxy}
\end{equation}
Here $h/e^2$ is the quantum of resistance, equals to about
$25.813{\rm k}\Omega$. The dimensionless number $\nu$ has, 
in the observed steps, either integer values $\nu=1,2,3...$
going up to several tens, or "simple" fractions $p/q$. Most of the
simple fractions are of odd denominators (one of the exceptions to
that rule will be later discussed), usually smaller than about
$15$. Of those the most prominent fractions are of the series
$\nu=p/(2p+1)$, with $p$ an integer. Generally speaking, the lower
is the temperature and the cleaner is the sample, the more steps
are observed. Integer values of $\nu$ go under the name Integer Quantum Hall Effect (IQHE) while fractional values go under the name Fractional Quantum Hall Effect (FQHE).

Let us pause to emphasize - these steps are astonishing. First,
they are amazingly precise. Eq. (\ref{quantumrxy}) is satisfied to
a level of one part in $10^9$! Second, their value is independent
of any properties of the material being measured. Rather, they are
determined by the ratio of two of the four universal constants,
the charge of the electron and Planck's constant.

At the same magnetic field in which $R_{xy}$ is on a step, the
longitudinal resistance $R_{xx}$ {\it vanishes}. Current flows without local dissipation of energy.
Ohm's law may then be
put in local terms, as a ratio between the gradient of the
electro-chemical potential $\bf E$ (loosely speaking, the electric
field) and the current density $\bf j$ as,
\begin{equation}
{\bf E}=\frac{h}{e^2\nu}{\hat z}\times{\bf j}
\label{qheresistivity}
\end{equation}
where both vectors lie in the $x-y$ plane. 
As the electric field and the current are perpendicular to one
another, the vanishing longitudinal resistivity (commonly denoted
by $\rho_{xx}$) implies also a vanishing longitudinal conductivity
$\sigma_{xx}$, and the quantized Hall resistivity
$\rho_{xy}=h/e^2\nu$ implies a quantized Hall conductivity
$\sigma_{xy}=e^2\nu/h$.

These universal values are the zero temperature limit of the
experimental observations. The deviations at finite temperatures
are a rich issue by itself, which we touch only briefly: at
least within a certain range of temperatures, the deviation of
$\sigma_{xx},\sigma_{xy}$ from their zero temperature value
follows an activation law, i.e., is proportional to
$e^{-\frac{T_0}{T}}$, with $T$ being the temperature and $T_0$ being a temperature scale that depends on many details. This
temperature dependence, which is familiar also from other
contexts, such as low temperature conductance of intrinsic
semi-conductors or low temperature dependence of the heat capacity
of super-conductors, indicates the existence of an energy gap
between the ground state and the first excited states. The
deviations from the activation law, which are observed at low
temperatures, probably indicate that the gap is not a "true" gap
(a finite range of energies at which there are no electronic
states) but rather a mobility gap (a finite range of energies at
which there are no extended electronic states).

Faced with these dramatic experimental observations, one naturally
asks "why does the effect happen?". We will address this question very partially later. At the moment, however, we are
more interested in a different question, namely, "given that the
effect is observed, what does it teach us?" As we will see, the
very observation of a fractional quantized Hall conductivity, a vanishing longitudinal conductivity and a mobility gap between the ground state and the first excited states, when combined with general
principles of physics, will force us to accept the notion of
Anyons.

But for going along that route, we first need to review the
Aharonov-Bohm effect \cite{ABeffect59}.

\section{The Aharonov-Bohm effect}
Let us start by digressing a bit, and reminding ourselves a few
basic facts on the physics of electrons in a magnetic field. High
school physics tells us that an electron in a magnetic field $\bf
B$ and an electric field $\bf E$ is subjected to a force
\begin{equation}
{\bf F}=e{\bf E}+\frac{e}{c}{\bf v}\times{\bf B} \label{lorenz}
\end{equation}
Undergraduate classical physics tells us that this force may be
derived out of an action. In particular, if we choose a gauge in
which the scalar potential vanishes, this force results from one
term in the action,
\begin{equation}
\frac{e}{c}\int dt {\bf v(t)}\cdot{{\bf A}({\bf x},t)}
\label{action}
\end{equation}
Here ${\bf A}$ is the vector potential whose various derivatives constitute the
electric and magnetic fields.

Quantum mechanics tells us that an action is not only a device to
generate an equation of motion. Rather, when divided by $\hbar$ it
gives the phase of the contribution of the trajectory
${\bf x}(t)$ to the propagator of the particle. If we look at the
case in which the vector potential is independent of time, we see
that the term (\ref{action}) becomes geometric, i.e., independent
of the velocity the trajectory is traversed by,
\begin{equation}
\frac{e}{c}\int dt {\bf v(t)}\cdot{{\bf A}({\bf
x})}=\frac{e}{c}\int d{\bf l} \cdot{{\bf A}({\bf x})}
\label{abgeometric}
\end{equation}
where the integral is taken along the trajectory. Gauge invariant
quantities always involve closed trajectories (loops). For those,
Stokes theorem tells us that the integral (\ref{abgeometric})
becomes $2\pi\Phi/\Phi_0$, where $\Phi$ is the flux enclosed in
the trajectory, and $\Phi_0\equiv hc/e$ is the flux quantum. This
is the Aharonov-Bohm phase.

A particularly important case is the case of the vector potential created by a
solenoid that cannot be penetrated. In that case, the trajectories
of an electron may be topologically classified by $n_w$, the
number of times they wind the solenoid. The phase that corresponds
to a trajectory is then $2\pi n_w\Phi_s/\Phi_0$, where $\Phi_s$ is
the flux enclosed by the solenoid. For all trajectories, then,
this phase is periodic in the flux, with the period being the flux
quantum $\Phi_0$.

\begin{figure}
\includegraphics[width= 0.6\linewidth,angle=-90]{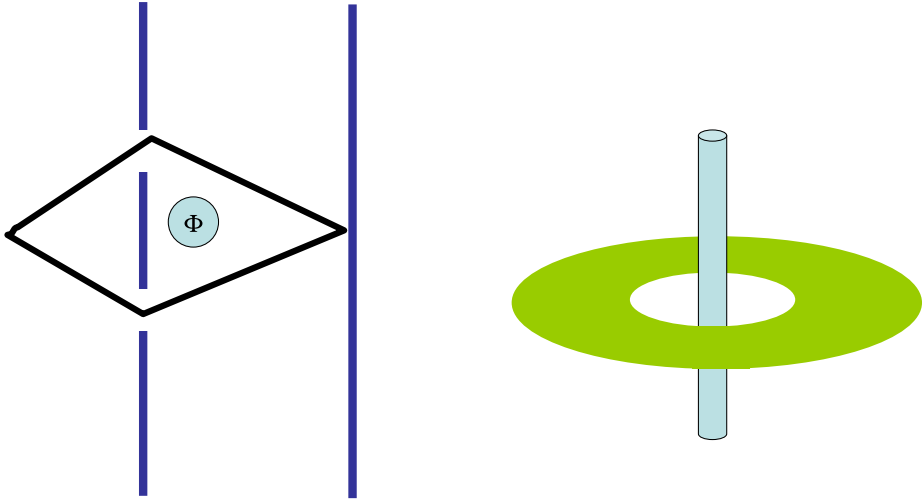}
\vspace{5mm}\caption{\label{ABeffect} The Aharonov-Bohm effect. Two realizations of the Aharonov-Bohm effect. In the first, shown in (a), the magnetic flux induces a shift in the interference pattern of a double slit experiment. An analog effect is to be seen in the quantum Hall interferometers discussed in Sections [\ref{interferometers-abelian}] and [\ref{interferometers-nonabelian}]. In the second, shown in (b), the magnetic flux affects the thermodynamic properties of the annulus around it. This realization is very useful in the analysis of the fractional charge and fractional statistics of quasi-particles in the quantum Hall effect, see Sec. [\ref{fractionalcharge}].}
\end{figure}

There are two useful experimental set-ups to think about in the
context of the Aharonov-Bohm effect (see Fig. \ref{ABeffect}). The first is the famous
double slit interference experiment. When a solenoid is put in
between the two slits, it shifts the interference patter due to
the phase shift it induces. Note that this shift in the
interference pattern is very surprising from a classical point of
view, as the electron, which cannot penetrate the solenoid, does
not feel any Lorenz force when it moves.

The second set-up is that of a ring, or an annulus, with a
magnetic flux $\Phi$ going through the hole. The electron is now
confined to the annulus and again does not experience any Lorenz
force due to the magnetic field in the solenoid. However, due to
the Aharonov-Bohm effect its  spectrum does depend on the magnetic
flux, with a period of $\Phi_0$. And since this is true for the
spectrum, it is true for all thermodynamic properties, that may
all be expressed as derivatives of the partition function.

The dependence of various thermodynamical quantities on the magnetic
flux through the hole, despite the absence of any force exerted on
the electron by the magnetic field in the solenoid, may be
understood in the following way: suppose that we look at a ring
where initially, at time $t=-\infty$, no flux penetrates the hole.
Then a flux $\Phi$ is turned on adiabatically in time.
Classically, the process of turning the flux on involves the
application of an electric field on the electron, and hence its acceleration. While there is a freedom of the position at which the electric field would operate, the condition of
adiabaticity implies that the time period at which the electric
field operates is much longer than the time it takes an electron
to encircle the ring, and thus the electron cannot avoid the effect of the field. As long as the electron is not
prevented by a trapping potential from encircling the ring, it
will experience a force due to the electric field, and will absorb
(kinetic) momentum and angular momentum. Due to the adiabaticity,
the electron stays in an eigen-state throughout the process in
which the flux is turned on. Thus, the eigen-state evolves in such
a way that it absorbs kinetic momentum and angular momentum.  As
long as this momentum is not fully transferred to static degrees
of freedom (impurities etc.), the eigenstate will have the
electron encircling the ring in a non-zero velocity. This is the source of the persistent current, or thermodynamic orbital magnetization, observed in normal mesoscopic rings\cite{Imrybook}.

There is a subtle point that needs to be understood in this argument: consider first an electron in a ballistic circular  ring, subjected to a magnetic flux that is adiabatically turned on. As we said, classically the electron is accelerated by the field. Its velocity is $\frac{e}{mc}\frac{\Phi}{2\pi L}$, with $L$ being the circumference of the ring and $m$ the electron mass. Quantum mechanically, if the electron was initially in the ground state it stays in an eigenstate, due to adiabaticity. It does not necessarily stay in the ground state, however. As the flux is being turned on, there are points of time in which the gap in the spectrum closes, and two eigenstates, one whose energy increases with flux and one whose energy decreases with flux, become degenerate in energy. Even when these states are degenerate, there is no matrix element that allows for a transition between them, and thus the electron does not go through such a transition. The reason for that is that the two states have different (canonical) angular momenta, and the rotational symmetry of the problem dictates a conservation of angular momentum. Any deviation from these perfect conditions, however, for example by the ring being imperfectly circular or imperfectly ballistic, introduces transitions between the two states, removes the degeneracy, and leads to the electron staying, in the adiabatic limit, in the lowest energy state. This subtle point will be of importance later.

\section{Fractionally charged excitations: combining the Aharonov-Bohm effect and the quantum Hall effect\label{fractionalcharge}}

With very little detailed modelling, we are now able to conclude, following Laughlin's original arguments, that a two dimensional system that shows the fractional quantum Hall effect must have quasi-particles that carry a fraction of an electron charge\cite{Laughlin83}. Following that, we will add in a few basic observations about the dynamics of vortices in fluids, and conclude that the fractional charge of the quasi-particles in the fractional quantum Hall effect implies also that the quasi-particles are anyons. While the anyonic nature of the quasi-particles was first realized by Halperin \cite{Halperin84}, based on the sequence of observed fractional quantum Hall states, the derivation we will follow here is closer to that of Arovas, Schrieffer and Wilczek \cite{Arovas84}.

\begin{figure}
\includegraphics[width= 0.6\linewidth,angle=-90]{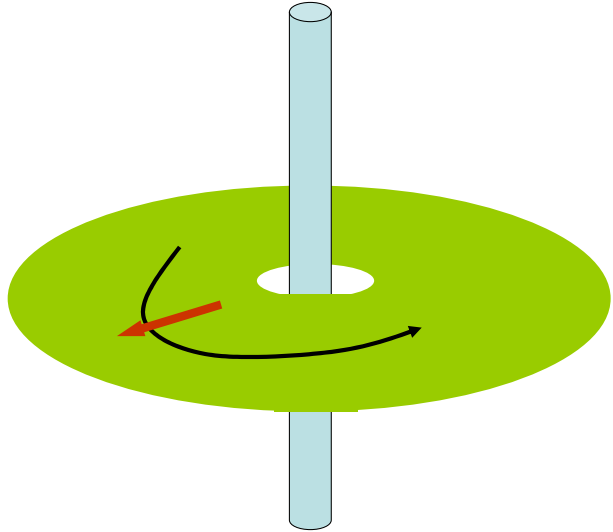}
\caption{\label{ABeffect-fractionalcharge} The gedanken experiment to create an eigenstate of a fractional charge, later realized to be an anyon. The electrons on the annulus are in a Laughlin fractional quantum Hall state of filling fraction $\nu=1/m$. The introduction of a flux quantum into the hole pushes a fraction of an electron charge from the interior to the exterior, leaving the system in a eigenstate. }
\end{figure}

The ingredients that will go into the heuristic derivation of the fractional charge are the Aharonov-Bohm effect, the fractional quantum Hall effect and the adiabatic theorem. Rather than looking at the ring we discussed before, let us consider a wide annulus, where the electrons on the annulus are in the (say) $\nu=1/3$ fractional quantum Hall state (see Fig. (\ref{ABeffect-fractionalcharge})). Again, we start with no flux threading the hole at the center of the annulus, and turn the flux on adiabatically. Again, an electric field is exerted on the electrons, and for simplicity we assume this electric field to be azimuthally symmetric. Let us analyze what happens when the flux is increased.

First, with the electrons being in a fractional quantum Hall state, the azimuthal electric field leads to a purely radial current flow, according to Eq. (\ref{qheresistivity}). At a distance $r$ from the origin the electric field is $E=\frac{1}{2\pi r c}\frac{\partial\Phi}{\partial t}{\hat \theta}$, and the resulting radial current density is $j_r=\frac{e^2\nu}{h}E$. Integrating the current density that goes through a circle of radius $r$ to get the total current, we find that the current does not depend on $r$. Thus, it leaves behind a charge lump at the interior edge of the annulus. Integrating the current over time, starting when there was no flux at the ring and ending at a time $t$ when the flux is $\Phi(t)$, we find that the amount of charge in that lump is
\begin{equation}
Q(t)=\frac{e^2\nu}{hc}\Phi(t)
\label{fraccharge}
\end{equation}

Second, in the quantum Hall effect there is an energy gap between the ground state and the first excited state. Thus, we can apply the adiabatic theorem, and conclude that the system, that started in a ground state when there was no flux, remains in an eigenstate throughout the process.

Third, what happens when the flux is $\Phi_0$? Then, Eq. (\ref{fraccharge}) gives a charge of $e\nu$. The system is at an eigenstate. The spectrum at $\Phi=\Phi_0$ is the same as the spectrum at $\Phi=0$ and the eigenstates at $\Phi=\Phi_0$ are the same as those at $\Phi=0$, up to the phase factor $\exp{i\sum_i\phi_i}$ (where $\phi_i$ is the azimuthal coordinate of the $i$'th electron), which does not affect the charge density.  Thus, we found that the annulus with $\Phi=0$ has an eigenstate in which a lump of charge of $e\nu$ is localized on its interior edge. The assignments of  energies to eigenstates may have interchanged, as in the case of a ballistic circular ring, so this is not necessarily the ground state. Clearly, an eigenstate in which a charge $e\nu$ has been transferred from the interior of the annulus to its exterior is physically different from the ground state the system started in. We call this state a quasi-particle when it carries a negative charge and a quasi-hole when it carries a positive charge. In much of the following the sign of the charge will not be of our concern, and we will use the terms quasi-hole and quasi-particle interchangeably.

As in the case of a ring, if the ground state at $\Phi=0$ evolves into an excited state at $\Phi=\Phi_0$ then for some intermediate value of $\Phi$ two states must have been degenerate in energy, with no transition between them taking place as the flux is varied. The absence of such a transition may be due to the absence of any matrix element connecting them (due to symmetries of the problem), or due to that matrix element being so small such that the rate at which $\Phi(t)$ would need to be changed becomes so slow to be impractical. Indeed, for the fractional quantum Hall annulus the latter is the case - the matrix element between the states is exponentially small in the size of the annulus (in the units of the magnetic length), and thus the two states are effectively uncoupled.

That this is the case may be understood in the following manner: the applied electric field acting in the azimuthal direction pushes a current to flow radially. If the system ends at $\Phi=\Phi_0$ in the initial state of $\Phi=0$, then  the charge that was driven from the interior to the exterior must have tunnelled back. This process of tunneling requires a breaking of rotational invariance, e.g., by impurities, but being a tunneling process, its amplitude is exponentially small\cite{DJTYG91}.

In the thought experiment we described above the quasi-particle/quasi-hole were excitations above the ground state. It is, however, easy to imagine the energetics to change in such a way that a state with such "excitations" becomes lower in energy than a state without them. That happens when electrostatic considerations force the density to deviate from a "magic" $\nu=1/m$ filling fraction. For example, imagine that rather than turning a flux quantum at the center of the annulus we would turn on a potential that would repel electrons away. If the potential is very small, it would have no effect, since the quantum Hall fluid is incompressible, due to its energy gap. But if the potential is strong enough it would lead to a ground state that includes one or more quasi-holes. Fractional charges of quasi-holes and quasi-particles have been observed in various types of measurements, such as resonant tunnelling through antidots \cite{Goldman95}, shot noise \cite{Picciotto97,Saminadayar97} and local compressibility \cite{Yacoby04}.

The argument we presented so far for the existence of fractionally charged quasi-particles and quasi-holes has been very general, being based on the existence of the fractional quantum Hall effect as a gapped ground state, and on general principles such as the adiabatic theorem and the Aharonov-Bohm effect. Historically, these quasi-particles were first understood through the use of trial wave functions. As Laughlin discovered\cite{Laughlin83}, if the ground state for a $\nu=1/3$ fractional quantum Hall state is assumed to be made solely of lowest Landau level wave functions, the following is a very good variational wave function for it (using complex coordinates $z_i=x_i+iy_i$ for the two Cartesian coordinates of the $i$'th particle):
\begin{equation}
\psi_{g.s.}(\{z_i\}, \{\bar z_i\})=\prod_{i<j}\left (z_i-z_j\right )^3\prod_i\exp{-|z_i|^2/4l_h^2}
\label{Laughlin}
\end{equation}
The virtues of this wave function will be explained later on (see Sec. [\ref{non-abelian-wf}]). At the moment we just say that it minimizes the kinetic energy by placing all the electrons in the lowest Landau level, and minimizes the potential energy by efficiently keeping electrons away from one another.
To generate a quasi-hole at the origin, Laughlin proposed the wave function
\begin{equation}
\prod_i z_i\psi_{g.s.}=\left(\prod_i|z_i|\right)e^{i\sum_i\phi_i}\psi_{g.s.}
\label{qhole}
\end{equation}
Indeed, the operator $\prod_i z_i$ does exactly what we expect the adiabatic introduction of a flux tube to do: the phase $\exp{i\sum_i\phi_i}$ gives each electron an angular momentum $\hbar$, while the product $\prod_i|z_i|$ moves each electron radially outwards, thus generating the radial current and the lump of charge at the interior. Over all, the operator does not introduce transitions between different Landau levels, and thus keep the wave function purely within the lowest Landau level.

The angular momentum carried by the quasi-hole (\ref{qhole}) is there not only for the Laughlin wave function of the quasi-hole. In fact, it is introduced also in the thought experiment that we described: as the flux is turned on, a radial current flows, with no introduction of angular momentum. However, when the wavefunction for $\Phi=\Phi_0$ is multiplied by $\exp{i\sum_i\phi_i}$ in order to get the wave function for the quasi-hole at $\Phi=0$, an angular momentum of $\hbar$ is given to every electron. The quasi-hole then carries a quantum of vorticity.

To summarize, the process we just described generates an eigenstate that carries a fractional charge $e\nu$ and {\it a single quantum of vorticity}, both localized at the interior edge of the annulus. By shrinking the size of the hole we can localize this charge to a size of the magnetic length (below that size the relation (\ref{qheresistivity}) does not hold). The state being an eigenstate implies that both the lump of charge and the vorticity do not decay in time. They are there to stay. 

\section{Charges and vortices - Lorenz and Magnus forces in  super-fluids and super-conductors}

The vorticity carried by the quasi-holes in the fractional quantum Hall effect will turn out to be crucial for their quantum statistics. Thus, let us remind ourselves of a few basic facts regarding the motion of vortices in two dimensions. In a Galilean invariant system, a vortex always moves at the speed of the fluid. In the absence of such symmetry, however, there may be a relative velocity between the vortex and the fluid. Under these conditions, Magnus force operates on the vortex. With this force being relevant for what comes next, we now review its basic features, and draw its similarity to the Lorenz force acting on charged particles in a magnetic field. We start by considering a vortex in a neutral superfluid, say superfluid Helium.

The origin of the Magnus force lies in the consideration of the vortex as a collective degree of freedom. Having a vortex centered at the point $\bf R$ implies that each fluid particle has an angular momentum $l\hat z$ relative to the point ${\bf R}$. When the vortex is static, there is a velocity field of the fluid, of ${\bf v}({\bf r})=\frac{l}{m}\frac{{\hat z}\times({\bf r}-{\bf R})}{|{\bf r}-{\bf R}|^2}$. When the vortex moves the velocity field of the fluid is the sum of this motion and the velocity of the vortex' center relative to the fluid, $\dot {\bf R}$. If the vortex' velocity $\dot{\bf R}$ is in the $\hat x$ direction, the velocity difference along the $\hat y$-axis leads to a pressure difference, and hence to a force in the $\hat y$ direction. The force is
\begin{equation}
F_{Mag}=2\pi nl {\hat z}\times {\hat v}
\label{magnus}
\end{equation}
where $n$ is the number density of the fluid. This force is
proportional to the product $nlv$ since the pressure difference
between the two sides of the vortex is proportional to $n\left (
({\bf v}+{\bf {\dot R}})^2-({\bf v}-{\bf{\dot R}})^2\right ) $.

Like the Lorenz force, the Magnus force is 
proportional and perpendicular to the velocity. Comparing Eq. (\ref{magnus}) to Eq. (\ref{lorenz}) we see that the role played by
the product $eB/c$ in Lorenz force is played by the product $2\pi
nl$ in Magnus force. As in the Lorenz force, the transition
from classical to quantum mechanics of vortices is best understood when
interference effects are considered. The classical Lorenz
force gives rise to the quantum mechanical Aharonov-Bohm phase accumulated by the wave function when an electron is moving, 
which then leads to the definition of the flux quantum. Similarly, the classical Magnus force gives rise in
quantum mechanics to a phase that is accumulated by the wave
function when a vortex is moving. The Aharonov-Bohm phase is
$\frac{e}{\hbar c}\int d{\bf s}\cdot {\bf B}$. The  analogous phase 
should then be
\begin{equation}
2\pi\frac{l}{\hbar}\int ds\  n
\label{vortexphase}
\end{equation}
For a vortex with a single quantum of vorticity (and all those we
are interested in are of this type) $l=\hbar$ and thus  the
phase accumulated by a vortex traversing a closed loop is $2\pi$
times the integral $\int ds\  n$ which is nothing but {\it $2\pi$ times
the number of fluid particles encircled by the vortex as it goes
around the loop.} The analog of the flux quantum is {\it a single fluid particle.}

We limited ourselves to a neutral fluid when we took the velocity
field of the vortex to decay as $1/|{\bf r}-{\bf R}|$, but, as we
now argue, both the classical force and the quantum mechanical
phase we obtained hold more generally. When the superfluid is
charged, as is a two-dimensional super-conductor, a vortex still
amounts to an angular momentum $\hbar$ given to each particle
fluid (in this case a Cooper pair), but the velocity field decays
as $1/|{\bf r}-{\bf R}|$ only for distances smaller than the London
penetration length, and decays exponentially for larger distance.
The reason for that difference is the magnetic field
created by the current that circulates around the center of the
vortex. The velocity of the $i$'th Cooper pair is
$\frac{1}{m}\left ({\bf p}_i-\frac{e}{c}{\bf A}({\bf r}_i)\right
)$, and at large distance $r$, the contribution of the canonical momentum ${\bf p}_i$ and the vector potential ${\bf A}({\bf r}_i)$ mutually
cancel. Despite this decay of the velocity field, the force that
acts on the vortex as it moves relative to the fluid does not
change. What changes is the source of the force: rather than being
the hydrodynamical Magnus force, it is now the Lorenz force. The
vortex is a solenoid of charge current enclosing a magnetic flux.
When it moves relative to a fixed background of charge, it exerts
an electric field on the charges, and a mutual force occurs. The
force is
\begin{equation}
\frac{1}{c}\int ds\left ( {\bf J}\times{\bf B}\right)\sim \frac{1}{c}2ne \frac{\Phi_0}{2}{\bf v}\times {\hat z}\sim 2\pi n l {\bf v}\times {\hat z}
\label{magnuslorenz}
\end{equation}
The first estimate is based on the current density being ${\bf
J}=2ne{\bf v}$ (the factor of $2$ originates from the charge of the
Cooper pair), the magnetic field  being ${\bf
B}\approx\Phi_0/2\lambda_L^2{\hat z}$ ($\lambda_L$ is the size
of the vortex, and the factor of $2$ originates from the flux of the vortex being half the flux quantum, $hc/2e$), and the integral being confined to a region of
$\lambda_L^2$.

 How is the
analog of the Aharonov-Bohm effect realized for vortices? It is
easiest to think of that in the context of a ring. As we saw, a flux
piercing the ring affects the spectrum of a particle residing on it, and
may induce a persistent current of the particle around the ring. Imagine now a vortex that is confined to move on a ring.
Such a confinement may be realized by a Josephson junction made of
two thin concentric super-conducting cylinders \cite{Elion93,Hermon94}, or by a two
dimensional super-conductor in which a circular trench is etched,
such that it is energetically costly for a vortex to reside out of
the trench. Now let us position the origin at the center of that
circle and imagine a radial current flowing into the circle. In
close analogy to the electron on a mesoscopic ring in which a
magnetic flux is adiabatically turned on, in the present case the radial current
exerts a force on the vortex and accelerates it. When the current
stops flowing the vortex is left with angular momentum, and keeps
on moving in a persistent way. The energy becomes a
function of the charge that flew from one side of the ring to
another, in quite the same way that it was a function of the flux
that was inserted to the ring in the previous problem.

A question now arises: the velocity field near the core of a vortex diverges,
and hence  the density there must be depleted so that the energy
stays finite. Since we found that the phase accumulated by a
vortex wave function as the vortex traverses a loop counts the
number of fluid particles enclosed in that loop, could there be a
topological phase associated with one vortex encircling another?
That is, could it be that the insertion of vortex $2$ into the
loop traversed by vortex $1$ would change the phase that the wave function accumulates by a fixed amount, preferably some
exotic fraction of $2\pi$?
We will now answer this question
negatively for vortices in neutral superfluids and
charged super-conductors and then proceed and see that the answer is
positive for quasi-holes in the quantum Hall effect.

In neutral super-fluids the density depletion at the core of the
vortex decays surprisingly slowly as a function of the distance
from the core. In fact, the decay is proportional to $r^{-2}$,
which leads to a diverging number of fluid particles pushed away from
the core of the vortex. Thus, no topological phase may
emerge\cite{HaldaneWu85}. Furthermore, both for neutral and charged superfluids,
there is no energy gap, as the collective charge density mode
(that may be called a plasmon, a phonon, or a Goldstone mode,
depending on taste) is gapless. Thus, when a vortex encircles 
another vortex in an infinite system, and when disorder removes the conservation of momentum, adiabaticity will never hold - the
process will necessarily involve a radiation of energy to that
mode, and that radiation will invalidate the calculation of the
phase.

Before we turn to the the quasi-particles and quasi-holes of the quantum Hall effect, we discuss
how our semi-heuristic arguments for the transition from the
classical Magnus force to the accumulation of a phase by the
vortex wave function can be made more rigorous. The main
ingredient we need for that is the geometric phase, better known
as Berry's phase\cite{Berry84}. Let us briefly review what that phase is and how
it comes into play in the context we deal with.

Generally, the systems we consider are composed of identical electrons whose coordinates are $\bf r_i$, and collective excitations, the vortices/quasi-particles/quasi-holes, whose coordinates are $\bf R_i$. We are interested in the effective Hamiltonians for the latter. That is, we are interested in turning the $\bf R_i$'s from parameters in the system's wave functions into dynamical degrees of freedom.  The vortex is a macroscopic degree of
freedom, so the dynamics of $\bf R$ is slower than the dynamics of
the fluid particles. In order to calculate the dynamics of $\bf R$ it is instructive to think of it as an external degree of freedom. For example, we think of the vortex as pinned to an impurity at the point $\bf R$, and the impurity as having a mass $M$ (this situation may naturally arise if the impurity repels the particles of the fluid. Since at the core of the vortex the density is depleted, the vortex would find it energetically favorable to be pinned on the impurity).  Note that we assume the impurity to interact only with the fluid particles, and not with any external fields exerted on the system. Any effect it feels of these fields must then arise from the effect of the fields on the fluid particles.

The question we ask is this: what is the effect of the vorticity carried around the point $\bf R$ on the dynamics of $\bf R$. The answer to that is given within an extension of the Born-Oppenheimer approximation. If the wave function of the system with a vortex at $\bf R$ is denoted by $|\psi({\bf R})\rangle$, then the kinetic energy that dictates the dynamics of $\bf R$ includes a Berry vector potential
\begin{equation}
A_B={\rm Im}\langle \psi({\bf R})|{\bf \nabla}_{\bf R}|\psi({\bf R})\rangle
\label{berrya}
\end{equation}
This vector potential depends on ${\bf R}$. When it has a curl with respect to ${\bf R}$, there is a Lorenz force, both in classical and in quantum mechanics. When the quantum mechanics of the vortex is considered, the vector potential leads to an Aharonov-Bohm like effect. The effective Hamiltonian of $R$ includes also an induced scalar potential, which is not of our concern here \cite{Berry84}. 

The precise wave function of the vortex depends on many details. Two basic properties are rather general, however: each fluid particle gets an angular momentum $\hbar$ relative to the point $\bf R$, and the density close to $\bf R$ must be depleted, to avoid a divergence of the kinetic energy. An operator that achieves both is
\begin{equation}
{\hat V}\equiv \prod_i e^{i\phi({\bf r}_i-{\bf R})}G(|{\bf r}_i-{\bf R}|)
\label{vortexoperator}
\end{equation}
with $\phi({\bf r})\equiv\arctan\left (\frac{y}{x}\right )$ being the angle conjugate to the angular momentum, and with the real function $G(r)$ vanishing at $r=0$. This operator creates a circularly symmetric vortex, but that simplification is not crucial. When it is applied to a featureless state, which is independent of $\bf R$, the geometric vector potential (\ref{berrya}) is, with $\rho({\bf r})$ being the fluid density at the point $\bf r$,
\begin{equation}
A_B({\bf R})=\int d{\bf r}\rho({\bf r})\frac{{\hat z}\times ({\bf r}-{\bf R})}{|{\bf r}-{\bf R}|^2}
\label{particlesaaretubes}
\end{equation}
This is precisely what our semi-heuristic analysis has led us to believe: the vortex, whose coordinate is $\bf R$, sees the fluid particles as an electron sees flux tubes. When it moves relative to the fluid, it experiences a force perpendicular and proportional to its velocity relative to the fluid, and the  magnitude of that force is exactly that given by the Magnus force. Eqs. (\ref{berrya}) and (\ref{particlesaaretubes}) also explain why the magnitude of the force is the same for neutral and charged vortices: it is determined by derivatives of the wave function with respect to ${\bf R}$, and the wave functions for vortices in neutral and super-conducting vortices share the same crucial property: the canonical angular momentum of $\hbar$ given to each fluid particle relative to the vortex' center. The magnetic field created by the currents that form the vortex in a super-conductor affects the kinetic angular momentum $m{\bf v}\times {\bf r}$ of the Cooper-pairs around the center of the vortex, but does not affect their {\it canonical} angular momentum ${\bf p}\times {\bf r}$. Finally, the phase accumulated by the quantum mechanical wave function when the vortex traverses a closed loop is $2\pi$ times the expectation value of the number of fluid particles enclosed in the loop \cite{Arovas84}.

\section{quasi-holes in the fractional quantum Hall effect follow fractional statistics}

So we finally got to the quasi-holes in the fractional quantum
Hall effect. As we saw, these quasi-holes, at the Laughlin
fractions of $\nu=1/m$, carry a fraction $1/m$ of the electron
charge and one quantum of vorticity. What can we say about their
dynamics, using the two other types of vortices as references?

 The phase factors in the operator (\ref{vortexoperator}) create the vorticity carried by the quasi-hole, and with the proper choice of $G(r)$ this operator will create also the correct depletion of charge. While there are many choices that would satisfy this condition, the most natural one, as suggested by Laughlin, is $G(r)=r$, making the quasi-hole operator $\prod_i z_i$ \cite{Laughlin83}. Under this choice, the operator that creates the quasi-hole does not introduce mixing of Landau levels. When it is applied to the Laughlin wave function (\ref{Laughlin}), the resulting quasi-hole state is still purely within the lowest Landau level.

Similar to the case of a vortex in a charged super-conductor, the velocity field of the vortex created by the quasi-hole creation operator does not decay inversely with the
distance from the center of the quasi-hole. Within the Laughlin picture, the charge density and azimuthal current associated with the quasi-hole both fall off exponentially at distances larger than the magnetic length. A more elaborate analysis \cite{Sondhi92} realizes that the azimuthal currents must be proportional to $1/r^2$ (with $r$ being the distance from the quasi-hole), since the charge of the quasi-hole creates a radial electric field that decays as $1/r^2$, and that field generates an azimuthal Hall current. Furthermore, the charge density is found to decay as $1/r^3$. Independent of these details,
however, as the quasi-hole moves relative to the quantum Hall
fluid, it experiences a force proportional and perpendicular to
its velocity, given by $2\pi n l{\bf v}\times {\hat z}$, resulting from the vector potential (\ref{particlesaaretubes}). This is
the same Magnus force experienced by a vortex in a neutral
super-fluid, and the same Lorenz force that the vortex in a
charged super-conductor experienced due to its interaction with the current. Here this
force has a rather simple description: noting that the angular
momentum is $\hbar$ and that $n\Phi_0/B=\nu$ we see that the force
acting on the quasi-hole is $\frac{e\nu}{c}{\bf v}\times{\bf B}$,
i.e., it is the Lorenz force acting on a charge $e\nu$ as it moves
in a magnetic field $\bf B$. Viewed as such, this is not too surprising - if we accept the quasi-hole as a collective degree of freedom that carries a fractional charge, it should be subjected to the Lorenz force that corresponds to that charge.

But the story gets more interesting when  we think of a quasi-hole as a quantum mechanical degree of freedom, i.e., when we consider the phase it accumulates due to the vector potential (\ref{particlesaaretubes}). The quantum Hall fluid is incompressible. Thus, when the density is such that the local filling factor deviates from $1/m$, the deviation is accommodated in integer number of quasi-particles or quasi-holes. Given the charge $e\nu$ of each quasi-hole, {\it when a quasi-hole goes around a closed loop, every quasi-hole it encircles contributes a phase of $2\pi\nu$ to its wave function. And since a winding of one quasi-hole around another is just two interchanges, we find that an interchange of the position of two quasi-holes multiplies the wave function by a phase of $\pi\nu$, which makes the quasi-holes anyons} \cite{Arovas84}.

Note again that what gives the quasi-holes their anyonic character is a combination of two ingredients - the fractional charge and the vorticity they carry. Both are essential for the quasi-hole to exist, as we described here in detail. 

\section{Interferometry as a way to observe anyons\label{interferometers-abelian}}

If anyons are defined by the quantum mechanical phase that they
accumulate when they encircle one another, it makes sense to
design experimental devices in which quantum mechanical phases are
observables and quasi-particles encircle one another. The
observation of phases accumulated by waves is the essence of
interferometry. We thus turn to interferometers as the devices for
the observation of anyons, dealing with the
Fabry-Perot and Mach-Zehnder interferometers.

\subsection{Fabry-Perot interferometer}

\begin{figure}
\includegraphics[width= 1.0\linewidth]{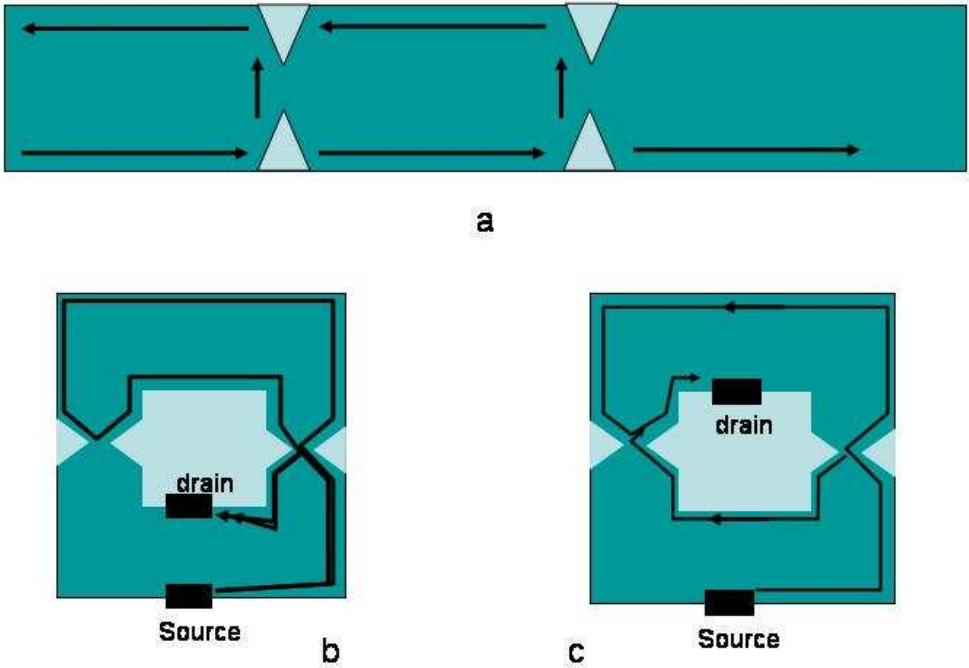}
\caption{\label{interferometers} The Fabry-Perot (a,b) and Mach-Zehnder (c) interferometers. The second drawing is meant to emphasize the difference between the two interferometers. The interior edge is a part of the interference loop in the Mach-Zehnder interferometer, while it is not part of that loop in the Fabry-Perot interferometer. Furthermore, in the former only single tunnelling events take place, while the latter allows for multiple reflections and the formation of resonances. }
\end{figure}

In the context of the quantum Hall effect, the Fabry-Perot
interferometer, discussed in details first by Chamon {\it et al.} \cite{Chamon}, is a Hall bar (for concreteness, lying along the
$x$-axis, between $0<y<w$) perturbed by two constrictions, known as quantum point contacts
(say, at $x=\pm d/2$). Current flows chirally along the edges,
towards the left along the $y=w$ edge and towards the right along
the $y=0$ edge. The right moving edge is put at a chemical
potential difference $V$ relative to the left moving one. In the
absence of the point contacts there is no back scattering of
current, and thus when the bulk is in a quantum Hall state the
two-terminal conductance (the ratio of the current difference to
the voltage difference between the two edges) is quantized. The
point contacts introduce an amplitude for inter-edge tunnelling,
which makes some of the current flowing along the right moving
edge back-scattered. 
The probability for back-scattering is
directly reflected in the voltage between the edges, i.e., in the
two terminal conductance.
Fig. (\ref{interferometers}b) folds the Hall bar into an annulus, for the purpose of comparison with the Mach-Zehnder interferometer, to be discussed later. In this setting, the point contacts introduce a flow of tunnelling from the source at the exterior to the drain in the interior. 

The presence of two point contacts makes the probability for
back-scattering the quantum Hall analog of "the interference
screen". If the tunnelling amplitudes introduced by the two point
contacts are $t_1,t_2$ respectively, then the back-scattered
current is, to lowest order in these amplitudes, proportional to
$|t_1+t_2|^2$. To this lowest order, the Fabry-Perot interferometer is the QHE analog of the two slit experiment. The relative phase between the two amplitudes may
be varied in three principal ways: by varying the magnetic field,
by varying the electron density and by varying the area of the
"cell" defined by the two edges and two point contacts. The latter
is implemented by a side gate that "moves the walls" of the cell
in the region it operates on. Within the crudest approximation, a
variation of the area of the cell does not vary the filling factor
in the bulk. Thus, it varies the number of electrons in the cell
(which is the product of density and area), but does not introduce
quasi-particles into the bulk. The variation of the electron
density and/or magnetic field lead to a more complicated effect.
Electrostatic interaction tends to fix the electronic charge
density to be such that it exactly neutralizes the positive
background. When this interaction dominates, a variation of the
bulk Landau filling from that of the quantum Hall state (say, a
variation from an integer $\nu$ or from $\nu=1/m$) leads to the
introduction of localized quasi-holes or quasi-particles. Such a
variation may be created either by varying the magnetic field or
by varying the positive background density (usually done by
applying a voltage to a gate). 

Given this picture, the Fabry-Perot interferometer gives us a
device in which quasi-holes that are flowing on the edge split
into partial waves that interfere, the interference pattern is
measured by measuring the backscattered current as a function of
three possible parameters, and the area enclosed by the
interfering trajectories enclose localized quasi-holes. It then
makes sense to expect the dependence of the back-scattered current
on the area of the cell, the magnetic field and the density to
reflect the statistics of the quasi-particles. Let us see how that
happens.

We start with the simplest toy model, non-interacting electrons
at $\nu=1$, and consider how the back-scattered current depends on
the area of the cell and the magnetic field. By the Aharonov-Bohm
effect, the relative phase between the two interfering waves is
$2\pi$ times the number of flux quanta enclosed in the cell
between the two interfering waves. That number is $BS/\Phi_0$,
with $S$ being the area of the cell. As $S$ is varied, then, we
expect a sinusoidal variation of the back-scattered current. The
period of this variation is $\Delta S=e/n_0$, with $n_0$ being the density
of electrons. In other words, the number of electrons in the cell
changes by one for each period of the oscillations.

As $B$ is varied the picture might be slightly more complicated,
since the variation of $B$ might also lead to a variation of $S$,
by changing the occupation of states at the edge. These edge
effects are important when the back-scattering probability is
close to one, such that the cell becomes a quantum dot. Deferring
this limit to a later stage, we assume the area of the cell to be
independent of the magnetic field, and thus the dependence of the
back-scattering current on the magnetic field to be sinusoidal as
well. The period for these oscillations is $\Delta B=\Phi_0/S$.

It is important to note that as the magnetic field varies and the
bulk filling factor becomes, say, $1-\epsilon$, a density $\epsilon n$
of quasi-holes, which for $\nu=1$ are just missing electrons,
appears in the bulk. On average, a variation of the magnetic flux
through the cell by one flux quantum introduces one quasi-hole
into the bulk of the cell. The precise values of magnetic field in
which a quasi-hole is introduced into the bulk depends on the
detailed geometry and the disorder potential within the cell.
While for $\nu=1$ the introduction of quasi-holes into the bulk
does not affect the back-scattered current, this is not the case
for the fractional quantum Hall states, as we now see.

When the filling factor is the fraction $\nu=1/m$ several factors
should be taken into account. First, several objects may be
tunnelling across the point contacts, starting from
quasi-particles of charge $e\nu$ and ending with electrons of
charge $e$. For several reasons, we expect the tunnelling of
$e\nu$-charged quasi-particles to be the strongest: being the
object with the smallest charge, its bare tunnelling matrix
elements are likely to be the largest. And under renormalization
at low temperatures and low voltage, it has the fastest growth
rate. We therefore focus on this type of tunnelling. Second, the
phase accumulated by a quasi-particle encircling the cell between
the two point contacts is
\begin{equation}
\phi=2\pi\nu\left [BS/\Phi_0-N_{qh}\right ] \label{fractionphase}
\end{equation}
This phase could be thought in two ways: it is the sum of the Aharonov-Bohm
phase accumulated by a charge $e\nu$ going around a flux $BS$ and
the statistical phase of a quasi-particle going around $N_{qh}$
others. Alternatively, it is the geometric phase we calculated
above, a phase of $2\pi$ per every fluid particle encircled. In
any case, when the area $S$ is varied, the period of the
oscillations of back-scattering is $\Delta S=e/n_0$, as before. When the
magnetic field is varied, on the other hand, there is a continuous
evolution of the Aharonov-Bohm part of the phase, accompanied by
occasional jumps of the statistical phase, taking place whenever
the number $N_{qh}$ increases by one. {\it These jumps are the
manifestations of the statistical interaction of the quasi-holes
flowing along the edge with those that are localized in the bulk.} 

When the amplitude for tunnelling across the two point contacts is
not very small, multiple reflections should be taken into account. Waves that are back-scattered at the second point-contact wind the cell several times before leaving the cell towards one of the contacts. The electronic Fabry-Perot interferometer develops resonances: rather than having the backscattered current oscillate sinusoidally as a function of magnetic field and area, most of the current is backscattered, except in certain lines in the magnetic field-area plane in which constructive interference of the waves that are transmitted through the cell build up the probability of the current not to be back-scattered. Generally, the transmitted current takes the form:
\begin{equation}
I_{bs}=\sum_{n=0}^{\infty}I_n \cos{n(\phi+\alpha_0)}
\label{multiplereflections}
\end{equation}
where the phase $\phi$ is given in (\ref{fractionphase}) and $\alpha_0$ is a phase that originates from the phases of the tunnelling amplitudes of the two point contacts. The phase $\alpha_0$ is independent of the magnetic field and the area. Resonances take place when  $\phi-\frac{2\pi N_{qh}}{m}=2\pi \ell$, with $\ell$ an integer. The area between the resonances is then $e/n_0$.

This state of affairs can be rephrased in a different language: in the limit of strong back-scattering the cell becomes a quantum dot, in which the number of electrons is quantized. For most values of the area and the magnetic field, the dot's energy is minimized with a unique number of electrons $N(S,B)$. Current through the dot is then blocked by the energy cost associated with adding the flowing electron to the dot. Resonances take place at those value of $S$ and $B$ for which the energies of the dot with $N$ electrons and with $N+1$ electrons are degenerate. One expects these degeneracy points to appear at an area separation of $e/n_0$. 

The picture we described is based on a sharp distinction between edge quasi-holes, which are flowing from one contact to another, and bulk quasi-holes, which are localized. If this distinction is realized experimentally, the Fabry-Perot interferometer takes quasi-holes to encircle one another in a way that realizes the gedanken experiments rather faithfully. Practically, however, it is difficult to avoid the number of quasi-particles enclosed in the cell, $N_{qh}$, to fluctuate during the experimental time scale. When such fluctuations take place on a characteristic time scale $\tau_0$ in which many quasi-holes are being back-scattered, they result in two effects: first, for a given value of $S$ and $B$ the back-scattered current becomes time dependent, and hence noisy. In the limit of weak backsttering the noise 
\begin{equation}
\left [\langle I_{bs}^2\rangle-\langle I_{bs}\rangle^2\right ]_{\omega=0}\propto \langle I_{bs}\rangle^2\tau_0
\label{oneoverftype}
\end{equation}

Second, as seen from (\ref{multiplereflections}), certain harmonics, those for which $n\nu$ is an integer, survive the averaging over temporal fluctuations of $N_{qh}$. The time averaged back-scattered current then becomes $\Phi_0$--periodic. Furthermore, in the limit of weak back-scattering, the Aharonov--Bohm oscillations of the back-scattered current would originate primarily from the $n\nu=1$ term in (\ref{multiplereflections}). Since 
\begin{equation}
I_m\propto I_0^{m-1},
\label{visibilityfp}
\end{equation}
the visibility of the Aharonov-Bohm oscillations, for a fixed voltage and varying back-scattering strength, would scale like the average current to the $m-1$ power.

\subsection{The Mach-Zehnder interferometer}

The Mach-Zehnder interferometer is another device that allows for
an interference of trajectories of particles that are
back-scattered through two point contacts. The two important differences between
the Fabry-Perot and Mach-Zehnder interferometers may be
conveniently viewed when they are depicted as in Fig.
(\ref{interferometers}). First, while the former allows for multiple reflections and the
formation of resonances, the latter interferes only two waves,
each one going through one tunnelling event. The structure is built in such a way that each quasi-hole may tunnel between edges at most once before being collected at the contact. Second, 
while in the former the interference loop encloses only the cell
between the two point contacts, in the latter the entire internal
edge is part of the interference loop. 

The inclusion of one of the edges in the interference loop has a profound effect on the interference pattern that the interferometer shows, as analyzed by Law {\it et al.}\cite{Law06, Feldman07} (see also \cite{ponomarenko:066803,Kane03}). The reason for that is that $N_{qh}$ now includes the number of quasi-holes on that edge, and this number changes by one with every tunnelling quasi-hole. Rather than discussing the back-scattered current directly it is then instructive to analyze the tunnelling rate of quasi-holes between the two edges. This rate depends on the number $N_{qh}$: 
\begin{equation}
\frac{1}{\tau}=\Gamma_0+\Gamma_1\cos{(\phi+\frac{2\pi N_{qh}}{m})}
\label{mzrates}
\end{equation}
Here $\Gamma_0$ is the classical term, of tunnelling either at the first or at the second point contact, and $\Gamma_1$ is the interference contribution. The latter has $m$ possible values, determined by $N_{qh}{\rm mod}\ m$. At any given time, then, the system is in a state in which, say, $j$ quasi-holes have tunnelled from the exterior edge to the interior one. The transition rates from that state to the state with $j\pm 1$ quasi-holes depend on $j{\rm mod} m$. A transition from the state $j$ to the state $j\pm 1$ implies a transfer of $\pm 1$ quasi-hole from one edge to another, i.e., a current. 

The flow of current through the interferometer is a statistical process, and the current average and noise may be calculated by means of rate equations. At zero temperature transitions are allowed only in one direction, determined by the difference of chemical potentials between the edges. The probability of the system being in a state where $j$ quasi-holes have tunnelled by time $t$, denoted by $P_j(t)$, then satisfies, 
\begin{equation}
\frac{dP_j(t)}{dt}=\frac{P_{j-1}}{\tau_{j-1}}-\frac{P_j}{\tau_j}
\label{rateequation}
\end{equation}
where $1/\tau_j$ is the rate for a transition from the state $j$ to the state $j+1$. The initial condition is 
$P_j(t=0)=\delta_{j,0}$.
The average current is $\langle I\rangle \left\langle e^*\frac{dj}{dt}\right\rangle=e^*\frac{\langle j(T)-j(0)\rangle}{T}$ for a very long time $T$, and the zero-frequency noise is ${\cal S}_{\omega=0}=(e^*)^2\left\langle\frac{dj}{dt}\frac{dj}{dt}\right\rangle_{\omega=0}$. Here $e^*=e/m$ is the charge of the tunnelling quasi-particle. A standard calculation shows that 
\begin{equation}
\left\langle\frac{dj}{dt}\right\rangle=\frac{m}{\sum_{i=1}^{m}\tau_i}
\label{currentmz}
\end{equation}
and 
\begin{equation}
\left\langle\frac{dj}{dt}\frac{dj}{dt}\right\rangle_{\omega=0}
=2m^2\frac{\sum_{i=1}^m\tau_i^2}{\left(\sum_{i=1}^m\tau_i\right )^3}
\label{mznoise}
\end{equation}
Substituting the rates Eq. (\ref{mzrates}) into these expressions, we find that both the average current and the current noise depend periodically on the flux, with a period of $\Phi_0$. The ratio ${\cal S}/2\langle I\rangle$, the Fano factor, that is frequently interpreted as the effective charge $q$, is
\begin{equation}
q=e\frac{\sum_{i=1}^m\tau_i^2}{\left(\sum_{i=1}^m\tau_i\right )^2} 
\label{fano-abelian}
\end{equation}
For $m=1$, where quasi-holes are fermions, the effective charge is just the electron charge. While both the current and the noise depend on the flux, their ratio does not. In contrast, for the Laughlin fractions, with an odd $m>1$, the effective charge generally depends on flux. This dependence is a direct consequence of the fractional statistics of the quasi-holes. It originates from the fact that the tunnelling rate of a quasi-hole from one edge to another depends periodically on the number of quasi-holes that have already tunnelled, and that dependence is a direct consequence of the geometric phase that a quasi-hole accumulates when it encircles another one. 

The effective charge in a Mach-Zehnder interferometer, Eq. (\ref{fano-abelian}) may be rather easily understood in two cases. In the case where all the tunnelling rates are the same, the effective charge is the charge of the quasi-particle, $e/m$. In the case where the rate to tunnel out of one state is significantly smaller than all others, the system will spend most of its time waiting to tunnel out of that state. Once it does so, it very quickly goes through the $m$ steps until it gets back to that state. In that case, then, the effective charge will be $m$ times larger than the quasi-particle charge, and that would amount to the charge of the electron. All other cases yield an effective charge between $e/m$ and $e$. As we will see later, values of the charge that are larger than $e$ are indicative of non-abelian anyons.

The Mach-Zehnder interferometer exhibits also a relation between the visibility of the Aharonov-Bohm oscillations, similar to that of (\ref{visibilityfp}), as verified by using Eqs. (\ref{currentmz}) and (\ref{mzrates}). The current depends on the flux through $I(\Phi)=I_0+I_1\cos{\left (2\pi \frac{\Phi}{\Phi_0}+\alpha_0\right )}$, with a power law relation between $I_1$ and $I_0$ \cite{Law06}. 

On a practical note it is worth mentioning that the increase of the visibility of the oscillations with increasing current, which we find both for a noisy Fabry-Perot interferometer and for a Mach-Zehnder interferometer, is opposite to the dependence one would expect from the obvious effect of heating. The latter increases with increased current, and suppresses the visibility due to loss of coherence. 

A different scheme for observing subtle signs of non-abelian statistics in current noise was suggested in \cite{Kim06}

\section{Anyons, composite fermions and the $\nu=p/(2p+1)$ states\label{cftheory}}

So far we dealt with anyons in the $\nu=1/m$ states, basing our analysis on little more than the experimental input of the fractional quantum Hall effect and general physics principles. In this section we show how anyons emerge from composite fermion theory, which is the most commonly used theoretical method to deal with the fractional quantum Hall effect\cite{Jain-book-07,Heinonen98,DasSarma97,Halperin93,Jain89,PhysRevB.47.7080,PhysRevB.44.5246}. There are some reasons to do that. First, the name composite fermion theory seems to suggest that one can understand the quantum Hall effect without ever worrying about fractional statistics. Here we explain how quasi-particles are anyons also in this theory. And second, composite fermion theory gives a theoretical tool for the treatment of fractional quantum Hall states not from the $\nu=1/m$ type. We will use this tool to analyze the statistics of quasi-particles in these states\cite{Blok90}.

So, in brief, what is composite fermion theory? The concept was first introduced by Jain in a first-quantized way, suitable for use in numerical work (\cite{Jain89} and see Section (\ref{non-abelian-wf})). In this Section we present it in a field theoretical way\cite{PhysRevB.47.7080,Halperin93}. The starting point is the well accepted Hamiltonian
\begin{equation}
H=\int d^2r\frac{1}{2m}\left |(i{\bf \nabla}-{\bf A})\psi(r)\right |^2+H_{\rm int}
\label{ham0}
\end{equation}
where $H_{\rm int}$ is the Coulomb interaction part of the Hamiltonian, and ${\bf A}({\bf r})=\frac{e}{2c}{\bf B}\times{\bf r}$. In this Hamiltonian $\psi({\bf r})$ is the electronic annihilation operator. The goal of the theory is to formulate a transformed Hamiltonian in which a natural approximation leads to the phenomenology of the quantum Hall effect, including an energy gap and a  quantization of the Hall resistivity.

The Chern-Simons transformation, which is the cornerstone of composite fermion theory, introduces the composite fermion annihilation operator
\begin{equation}
\psi_{\rm cf}({\bf r})=\psi({\bf r})e^{2i\int d^2r'\rho({\bf r}'){\rm arg}({\bf r}-{\bf r'})}
\label{cstrans}
\end{equation}
where $\rho({\bf r})=\psi^\dagger({\bf r})\psi({\bf r})=\psi_{\rm cf}^\dagger({\bf r})\psi_{\rm cf}({\bf r})$ is the density operator. In terms of this operator the kinetic part of the Hamiltonian acquires a new vector potential and becomes, 
\begin{equation}
H=\int d^2r\frac{1}{2m}\left |(i{\bf \nabla}-{\bf A}+{\bf a})\psi(r)\right |^2+H_{\rm int}
\label{ham1}
\end{equation}
with ${\bf \nabla}\times{\bf a}=2\Phi_0\rho({\bf r})$. Figuratively, the transformation attaches two flux quanta to each electron, transforming it to a composite particle that carries a charge $e$ and two flux quanta, and follows fermionic statistics. While this description is correct, it should be used with caution, as we will shortly see.

On the face of it, the introduction of this vector potential just makes the problem harder, since on top of their electrostatic interaction, the particles now interact also with the flux tubes of one another. However, the transformation also opens the way for the desired approximation, a Hartree mean field approximation. Replacing the dynamical vector potential ${\bf a}({\bf r})$ by its expectation value maps the problem of electrons at a magnetic field $B$ to composite fermions at a magnetic field $b=B-2\Phi_0n$, with $n$ the average density. That mapping also introduces a correspondence between Landau level filling factors: the electronic filling factor $\nu_e$ translates to a composite fermion filling factor $\nu_{\rm cf}$ according to $\nu_e^{-1}=\nu_{\rm cf}^{-1}+2$. In particular, the prominently observed series of fractional quantum Hall states at filling fractions of $\nu_e=\frac{p}{2p+1}$ maps onto an {\it integer} quantum Hall effect for the composite fermions, with $\nu_{\rm cf}=p$ (Note that the number $2$ in the exponent in Eq. (\ref{cstrans}) may be replaced by any other even number for the description of states that do not belong to this series, such as $\nu=1/5$ or $\nu=2/3$). The energy gap then becomes a natural consequence of the filling of an integer number of composite fermions Landau levels. The Hall resistivity of the composite fermions is quantized to be $h/pe^2$. Since a current of the composite fermions involves also a motion of their flux tubes, which creates a Chern-Simons transverse electric field, the resistivity of the electrons and that of the composite fermions are related by $\rho_{xx}^e=\rho_{xx}^{\rm cf}$ and $\rho_{xy}^e=\rho_{xy}^{\rm cf}+2h/e^2$. This relation reproduces the correct $\rho_{xy}^e=h/e^2\nu$ value. 

With the featureless fractional quantum Hall liquid of $\nu=p/(2p+1)$ being described as $p$ filled Landau levels of composite fermions, what would be the quasi-particles and quasi-holes? The procedure we described above for creating a quasi-hole in a $\nu=1/m$ state, by the adiabatic turning on of a $\Phi_0$ flux tube, can be applied here as well, and will create a quasi-hole with a charge of $e\nu=ep/(2p+1)$. It is possible, however, to create a quasi-hole with a smaller charge. That should come as no surprise: In the $\nu=n$ integer quantum Hall effect, in a model where electron-electron interaction is neglected, the charge of the quasi-particle is clearly the charge of the electron. It is created by adding an electron to an otherwise empty Landau level, or - for a quasi-hole - by removing an electron from an otherwise full level. In contrast, the introduction of a flux quantum expels a charge larger by a factor of $n$. 

To identify the charge of the quasi-hole or quasi-particle for the $\nu=p/(2p+1)$ states we repeat the same process, for composite fermions. Let us imagine adiabatically annihilating a composite fermion at the origin, at the lowest angular momentum state of one of the filled Landau levels. This process of annihilation is done in two steps. First, the charge is taken out, a charge of $e$. Then, a flux tube carrying $2\Phi_0$ parallel to the external magnetic field is introduced. Its adiabatic introduction introduces an azimuthal electric field, which drives away a charge of $2pe/(2p+1)$, leaving a net charge of magnitude $-e/(2p+1)$ as the charge of the quasi-hole. 

This process may look confusing at first sight: if the charge of a composite fermion is the charge of the electron, how come that by annihilating a composite fermion we create a lump of a fractional charge? The answer is that  by adiabatically annihilating a composite fermion in one Landau level we also affect the wave functions of composite fermions in the other Landau levels. Let us look at this carefully: the annihilation of the composite fermion takes away a charge $q$ (which we are now set to determine again). Thus, it also varies the mean-field effective magnetic field $b=B-2\Phi_0n$ in the region from which the charge was taken. Since a charge is taken out, $b$ grows, and the magnetic flux in the region grows by $2\Phi_0q$, making the charge in each of the filled Landau levels grow by $2q$, since a filled Landau level has a fermion per flux quantum. Altogether, then, one Landau level lost the composite fermion that was taken out, but got an extra charge $2q$ because of the magnetic flux growing. The other $p-1$ levels got a charge of $2q$ each. The total charge then satisfies 
\begin{equation}
q=1-2q-2q(p-1)=1-2pq
\label{cfcharge}
\end{equation}
leading to $q=1/(2p+1)$.

As we saw for the $\nu=1/m$ states, the statistics of the quasi-particles is closely linked with the vorticity they carry. In the process we described above, in which a composite fermion is taken out of one of the filled Landau levels, the Landau level from which it is taken out acquires a vorticity in exactly the same way that the quasi-hole introduced a vortex into the $\nu=1/m$ state. The other Landau levels, however, do not acquire any vorticity. When the quasi-particle goes around a closed trajectory, a phase is accumulated by the vortex that encircles a charge, but the the charge is only the charge in the Landau level in which the vortex resides. Consequently, if we consider two quasi-holes, one created by taking a composite fermions off the filled $i$'th Landau level, and the other created by taking a quasi-hole off the filled $j$'th Landau level, then the phase accumulated when one encircles the other is 
\begin{eqnarray}
2\pi(1-2q) &= 2\pi \left (1-\frac{2}{2p+1}\right ) &\  {\rm  when  }\  i=j\\
2\pi(-2q)&=-4\pi / (2p+1)&\ {\rm when  } \ i\ne j
\label{stat-p}
\end{eqnarray}

\section{From theory to experiments - present status and inherent difficulties}

There has recently been a surge of experimental activity in the field of mesoscopic quantum Hall devices in general, and interferometry in particular (see for example \cite{Camino, Camino05,Camino06,Camino07,godfrey-2007,miller07, willett-2007,neder:016804,ji03}).  Interesting and only partially explained phenomena were seen in Mach-Zehnder interferometers of integer filling factors \cite{neder:016804}, but those presumably  do not involve anyons. In a series of beautiful experiments of Camino {\it et al.}, devices of the Fabry-Perot type were fabricated, and were measured in the integer and fractional quantum Hall regime. The results of these experiments are not yet fully understood, and several interpretations have been put forward in subsequent theoretical works\cite{Averin07,Rosenow07a,Kim06,jain:136802,PhysRevB.75.045334,FieteRefael07}. While we will not get into a detailed discussion of these experiments here, we will describe the main results and comment on several factors that are crucial for their interpretation.  

Naturally, the place to start is with the integer quantum Hall effect. The experiment measured the dependence of the back-scattered current on the magnetic field and the electronic density, which was varied by means of a back-gate. Both dependencies were oscillatory. The period of the oscillations with respect to the back-gate voltage was independent of the integer filling factor $f$ at which the measurement was carried out. In contrast, the period with respect to the magnetic field was inversely proportional to $f$.

These measurements reveal a major difference between the theoretical construct we introduced above and its experimental realization: unlike in the theoretical construct, the "cell" of the interferometer is confined by a smooth potential, and therefore may break into several regions of different phases. One example to that is the center of the cell being an "island" of a  quantum Hall state of one filling factor surrounded by a quantum Hall state of a different (typically lower) filling factor. Another example is one in which the center of the cell is a compressible island, surrounded by a quantum Hall state. Yet another is one in which the bulk both outside and inside the cell is in one quantized Hall state but the region of the point contacts is in another. In all cases, the bulk of the cell has a compressible region, either as an edge separating an island of one quantized state from a bulk of another, or as a compressible  island within a quantized region. These compressible regions complicate the experiment in several ways. First, being confined by insulating quantized Hall regions, their charge is quantized. Second, they add  indirect paths for tunnelling from one edge to another. And third, their size is a degree of freedom that may vary as the magnetic field or back-gate voltage are varied.

Both the flux and the charge periodicity of the back-scattered current in the integer quantum Hall regime are understood in terms of Coulomb blockade physics of the compressible region within the interferometer's cell. As the flux within the island is varied by $1/f$ flux quanta, the occupation of the highest occupied Landau level (the one that is occupied only within the island) changes by one electron, hence giving rise to a Coulomb blockade periodicity of $1/f$. The periodicity in gate voltage is then naturally independent of the filling factor. 

Further measurements were carried out in the fractional regime, when the filling fraction at the constrictions was $\nu=1/3$ and that at the island between the constrictions was $\nu=2/5$. Oscillations of the conductance were observed again, with a flux period of five flux quanta through the estimated area of the $\nu=2/5$ island and a charge period of two electronic charges. It must be noted that the  estimate of the size of the $\nu=2/5$ island has a certain dependence on modelling, and is not directly measured. In any case, the anyonic nature of the quasi-particles is probably involved in the determination of the periods of the oscillations, but the precise way, and the role of the other factors, are not fully understood yet. In fact, different models of the experimental system yield different periodicities \cite{jain:136802,Rosenow07a,PhysRevB.75.045334}, none of which is presently able to fully account for the experimental observations. The models differ from one another in their identification of the dominant inter-edge tunnelling route and by the way they account for the relative roles of Coulomb blockade physics and interference effects.

\section{Non-abelian anyons \label{non-abelian}}
\subsection{General description}

So far we reviewed how the two-dimensional fractional quantum Hall effect extends the notion of quantum statistics and introduces anyons, particles whose interchange phases are between the zero of bosons and $\pi$ of fermions. From now on we will present the next level of extension, of particles satisfying non-abelian statistics, or non-abelian anyons\cite{RMP,MR,Blok92,Lo93,Wen91a,Frohlich89,Frohlich90,Wen98,Wen99,Bais80}. The first quantum Hall state suspected of being  non-abelian is the $\nu=5/2$ state\cite{MR,Willett87}. The study of this state has then led to the introduction of the Read-Rezayi series, the series of spin polarized $\nu=2+k/(k+2)$ states, for which non-abelian theories have been proposed \cite{Read99}), and to other non-abelian states, including states that are spin-singlets\cite{AS99, Ardonne2001,Simon07a}. Numerical works  indicate that the $\nu=5/2$ and $\nu=12/5$ ground state are indeed non-abelian for a wide range of interaction parameters \cite{Morf98,Morf03b,Rezayi06,Scarola02,Wan06}. 

We will start this Section with the $\nu=5/2$ state, both because of its relative simplicity and because of it being the most relevant to present days experiments, and will continue with the way Conformal Field Theories are used to reason and analyze the more complicated non-abelian states, taking the Read-Rezayi series as an example. 

Let us start by expanding on the statement we made regarding the effect of the quasi-hole's vorticity on the dynamics of its coordinate. As we said, if the wave function of the system with a quasi-hole at a coordinate ${\bf R}$ is $|\psi({\bf R})\rangle$, then the kinetic energy that dictates the dynamics of $\bf R$ includes a Berry vector potential
\begin{equation}
{\bf A_B}={\rm Im}\langle \psi({\bf R})|{\bf \nabla}_{\bf R}|\psi({\bf R})\rangle
\label{berryahitsagain}
\end{equation}
Then, when the vortex traverses a closed loop the wave function accumulates a Berry phase of $\int {\bf A_B}\cdot {d{\bf l}}$. 

There was an assumption in this statement, and its violation is the source of non-abelian quasi-particles. The assumption was that once the positions of the quasi-holes are fixed, there is just one ground state to the system. Under this assumption, when the parameters in that ground state, the positions $\bf R$'s, adiabatically traverse a closed loop, the state of the system must evolve from its initial ground state $|\psi({\bf R})\rangle$ to a final state that differs from the original one by a phase factor only. This is the adiabatic theorem. A calculation then shows that the phase factor is just the Berry phase mentioned above. 

What if the assumption does not hold, and rather than having one ground state per each configuration of quasi-holes we have several ground states that differ by some internal quantum numbers? Such a degeneracy may in principle appear accidentally, but here we are interested in the case where the degeneracy originates from deep properties of the quantum states. 

At any rate, with such a degeneracy, if the initial state is a ground state the adiabatic theorem guarantees that the final state will be a ground state as well, but it does not guarantee that it would be the initial state multiplied by a phase factor. Rather, under these conditions, the geometric vector potential ${\bf A}({\bf R})$ becomes a matrix, 
\begin{equation}
{\bf A_B}_{ij}={\rm Im}\langle \psi_i({\bf R})|{\bf \nabla}_{\bf R}|\psi_j({\bf R})\rangle
\label{berryamatrix}
\end{equation}
with the $|\psi_i\rangle$ being the various degenerate ground states. Then, the effect of a motion of the parameters ${\bf R}$'s along a closed loop is not a phase factor, but rather a unitary transformation that acts within the subspace of degenerate ground states. This unitary transformation is, 
\begin{equation}
{\hat P}e^{i\int{\bf A_B}\cdot {d\bf l}}
\label{unitarytrans-general}
\end{equation}
with $\hat P$ the path ordering operator. 

A degenerate subspace of ground states is just one condition for non-abelian anyons to occur. The second condition is that, up to an abelian phase,  the unitary transformation (\ref{unitarytrans-general}) will depend  only on the topology of the closed trajectory taken by the parameters $\bf R$. When that happens an interchange of particles is associated with a unitary transformation. Different unitary transformations do not necessarily commute and the state of the system after a series of interchanges of positions of particles may depend on the order at which these interchanges were carried out, hence the name non-abelian statistics. 

\subsection{The $\nu=5/2$ state}

This description was very abstract, and an example would probably help to clarify the idea. The most prominent example for non-abelian anyons are believed to be the quasi-holes and quasi-particles of the $\nu=5/2$ fractional quantum Hall state, as realized first by Moore and Read\cite{MR}. This state is also the easiest to analyze, using its mapping onto a $p$-wave super-conductor of composite fermions \cite{Read00,Greiter91, Greiter92,PhysRevB.54.16864}. Let us see how this comes about, following closely the work of Read and Green \cite{Read00}: 

In the $\nu=5/2$ state two Landau levels are full and one is half full. The two full levels are inert and for our discussion have no effect. The half filled Landau level may be mapped, by means of the Chern-Simons transformation (\ref{cstrans}), onto a system of composite fermions at an average of zero magnetic field, since the external field is cancelled by the Chern-Simons field. At very low temperatures it then becomes energetically favorable for the composite fermions, at least in some regime of interaction parameters, to form Cooper-pairs and have them condense to form a super-conductor.  If the electrons are assumed to be spin polarized (which is definitely the case in the limit of infinite magnetic field and density, and perhaps also for the experimentally relevant values) so are also the composite fermions. Then they cannot form $s$-wave Cooper-pairs, and the simplest pairing they may undergo is that of $p$-wave symmetry. In the absence of a spontaneous breaking of rotational symmetry, the two possible $p$-wave types of pairing would be $p_x\pm ip_y$.  To give an intermediate summary of  this line of argument, then, the $\nu=5/2$ quantum Hall state is to be thought of as a $p_x\pm ip_y$ super-conductor of composite fermions. 

With that in mind, note the important difference between the effect of a slight increase of the magnetic field (or a slight decrease of the density) on the $\nu=p/(2p+1)$ and the $\nu=5/2$ states: while for the former this change introduces vacancies in an otherwise full Landau level, for the $\nu=5/2$ the residual magnetic field is accommodated by the super-conductor in the form of vortices. These  vortices are the non-abelian quasi-holes and quasi-particles of the $\nu=5/2$ state, as we now see. 

As seen in Eq. (\ref{ham1}), the composite fermions are subjected to the vector potential ${\bf A}-{\bf a}$. The electromagnetic part of that vector potential is externally applied. The tiny currents that flow in a two dimensional electronic system do not produce a significant magnetic field. In contrast, the Chern-Simons vector potential {\it is} dynamical, i.e., created by the system itself. As in a super-conductor the vortex currents are screened by one half of a flux quantum of the gauge field ${\bf a}$. Since two flux quanta correspond to one electron charge, half a flux quantum corresponds to one quarter of a charge. The quasi-hole/quasi-particle are quarter charged, then. 

But this is not the end of the story. The next piece of it is the degeneracy of the ground state in the presence of these quasi-holes. To see that we need to analyze the $p_x+ ip_y$-wave super-conductor in some more detail (the choice of the relative sign is motivated by the need to obtain a ground state wave function of the lowest Landau level). The most straight forward way of analyzing a super-conductor is through the BCS mean field Hamiltonian, 
\begin{eqnarray}
&H&=\int \mathrm{d}{\bf r}\,\psi ^{\dagger }({\bf r})h_{0}\psi ({\bf
r})\nonumber \\ &+& {\frac{1}{2}} \int \mathrm{d}{\bf r}\,\mathrm{d}{\bf
r^{\prime }}\left\{ D^{\ast }({\bf r},{\bf r^ {\prime
}})\psi ({\bf r^{\prime }})\psi ({\bf r})+D({\bf r},{\bf
r^{\prime }})\psi ^ {\dagger }({\bf r})\psi ^{\dagger }({\bf
r^{\prime }} )\right\}  \label{bcs-hamiltonian}
\end{eqnarray}
with the single-particle term $h_0$ and the complex $p$-wave pairing
function
\begin{equation}
D({\bf r},{\bf r}^{\prime })=\Delta \left( {\frac{{\bf r}+{\bf r^
{\prime
}}}{2}}\right) (i\partial _{x^{\prime }}-\partial _{y^{\prime }})\delta
({\bf r}-{\bf
r^{\prime }}).
\end{equation}

The dynamics of $\Delta$ is governed by a Landau-Ginzburg-type
Hamiltonian. The quadratic
Hamiltonian (\ref{bcs-hamiltonian}) may be diagonalized by solving
the corresponding Bogolubov-de-Gennes equations, and following its diagonalization may be written as
\begin{equation}
H=E_0+\sum_{E>0} E\Gamma_E^\dagger\Gamma_E
\label{GammaH}
\end{equation}
where $\Gamma_E^\dagger\equiv\int dr \left [ u_E({\bf r})\psi({\bf r})
+v_E({\bf r})\psi^\dagger({\bf r})\right ]$ is the creation operator
formed by the positive energy solutions of the Bogolubov deGennes  (BdG)
equations,
\begin{eqnarray}
E\left(
\begin{array}{c}
u({\bf r}) \\
v({\bf r})
\end{array}
\right)  =\left(
\begin{array}{cc}
-\mu ({\bf r}) & \frac{i}{2}\left\{ \Delta ({\bf r}),\partial _{x}
+i\partial
_{y}\right\} \\
\frac{i}{2}\left\{ \Delta ^{\ast }({\bf r}),\partial _{x}-i\partial
_{y}\right\} & \mu ({\bf r})
\end{array}
\right) \left(
\begin{array}{c}
u({\bf r}) \\
v({\bf r})
\end{array}
\right),\label{BdG}
\end{eqnarray} 
and $E_0$ is the ground state energy. 
As becomes clear from Eq. (\ref{GammaH}), for the ground state of the Hamiltonian (\ref{bcs-hamiltonian}) to be degenerate it is essential that the BdG equations have zero eigenvalues. 

Such zero energy eigenvalues appear when the super-conductor has vortices.
Vortices are introduced into the Bogolubov-de-Gennes equations through $\Delta$. A vortex at the point ${\bf R}$ implies a winding of the phase of $\Delta$ by $2\pi$ for every trajectory that encircles $\bf R$, and a vanishing of $\Delta$ at the point ${\bf R}$ itself. Assuming, for simplicity, azimuthal symmetry, $\Delta({\bf r})=|\Delta(|{\bf r}-{\bf R}|)|e^{i\theta+i\Omega}$, where $\theta$ is the angle of the vector ${\bf r}-{\bf R}$ relative to the $x$-axis, and $\Omega$ is the phase of the order parameter along the $\theta=0$ line. 

When a vortex configuration for $\Delta$ is considered, and when the chemical potential $\mu$ is assumed larger than zero, the Bogolubov-de-Gennes equations are found to possess a single zero energy solution, localized close to the point ${\bf R}$, of the form
\begin{equation}
\gamma=\frac{1}{\sqrt{2}}\int dr \, \left [F
({\bf
r})\,e^{-\frac{i}{2}\Omega}\psi({\bf r})+\, F^*
({\bf
r})\,e^{\frac{i}{2}\Omega}\psi^\dagger ({\bf r})\right ]
\label{zeroenergy}
\end{equation}
Here, $F({\bf r})$ decays for large $r$.

When there are several well separated vortices at positions ${\bf R} _i$, the gap
function near the $i$'th vortex takes the
form $\Delta({\bf r}) = |\Delta({\bf r})| \exp{(i\theta_i+i\Omega_i)}$, with
$\theta_i=\arg{({\bf r}-{\bf R_i})}$ and $\Omega_i=\sum_{j\ne i}\arg(({\bf R_j}-{\bf R}_i))$. There is one zero energy solution per 
vortex, and we correspondingly add an index to $\gamma$. Each zero energy  solution $\gamma_i$ is localized near the core of its vortex at ${\bf R_i}$, but the  phase $\Omega_i$ that replaces $\Omega$ in (\ref{zeroenergy}) depends on the position of all vortices. Moreover, the dependence of $\gamma_i$ on the positions ${\bf R_i}$ is not single valued. If the angle of the vector ${\bf R}_i-{\bf R}_j$ with respect to the $x$-axis is changed by $2\pi$ then $\gamma_i$ acquires a minus sign. These minus signs will play an important role later.

The operators $\gamma_i$ have a few properties that we need to dwell on. First, as is obvious from (\ref{zeroenergy}), these operators are hermitian, $\gamma_i=\gamma_i^\dagger$. Second, they mutually anti-commute. With the proper normalization, $\{\gamma_i,\gamma_j\}=2\delta_{ij}$. These two properties make the $\gamma_i$'s Majorana fermions. Since $\gamma_i^2\ne 0$ and $\left (\gamma_i^\dagger\right )^2 \ne 0$, we cannot talk about the $\gamma_i$'s as being "empty" or "full" as we are so used to when talking about fermions. We need to be more careful when counting the dimensionality of the Fock space that is spanned by these fermions. To that end, we first note that the number of zero energy states must be even: for a compact geometry, Dirac's quantization of the monopole enforces the number of vortices going through the boundary-less surface to be even. For a non-compact geometry, if the bulk of the system has an odd number of vortices, the edge will have a zero mode. If the bulk has an even number of vortices, the edge will not have one of its own\cite{Read00}. Thus, there is always an even number of $\gamma_i$'s, with $i=1..2N$. To properly count the dimensionality of the Hilbert space they span, it is convenient to define complex fermionic operators, 
\begin{eqnarray}
\Gamma_j\equiv\gamma_j-i\gamma_{j+N}\\
\Gamma_j^\dagger\equiv \gamma_j+i\gamma_{j+N}
\label{majoranatocomples}
\end{eqnarray}
with $j=1..N$. The $N$ pairs of fermionic operators $\Gamma_j,\Gamma_j^\dagger$ satisfy the anticommutation relations we are used to from "conventional" fermionic operators, and hence define $N$ fermionic modes. In particular, each of these modes satisfy $\Gamma_j^2=\left (\Gamma_j^\dagger \right )^2=0$, and hence may be referred to as "empty" and "full". The dimension of the Hilbert space they span is therefore $2^N$, and a basis that spans it can be described in terms of binary numbers of $N$ digits, with $0$ denoting an "empty" mode, and $1$ denoting a "full" one.  Note that the BCS mean field Hamiltonian does not commute with the particle number operator, but does commute with the parity of that number. Thus, its eigenstates may be characterized by the parity of the number of particles they contain. Since the operators $\Gamma_j$ and $\Gamma_j^\dagger$ change that parity (each is a superposition of creating and annihilating a particle), half of the $2^N$ ground states have an even number of particles, and the other half have an odd number of particles.

This subspace of ground states is an interesting Hilbert space. The basis we chose is obviously arbitrary, being based on an arbitrary enumeration of the vortices. A different enumeration is just a different basis spanning the same subspace. But note how unique these states are: suppose that we look at a ground  state $|\psi\rangle$ where all fermionic modes are "empty", $\Gamma_j|\psi\rangle=0$ for all $j$. From Eq. (\ref{majoranatocomples}) we find that, 
\begin{equation}
\gamma_j|\psi\rangle=i\gamma_{j+N}|\psi\rangle
\label{nonlocal}
\end{equation}
Noting the structure (\ref{zeroenergy}) of the $\gamma$ operators, this relation tells us something interesting about the ground state $|\psi\rangle$. The vortices $j$ and $j+N$ may be arbitrarily far from one another. Yet, when we operate on $|\psi\rangle$  with $\gamma_j$, which is an operator that creates and annihilates a particle at the neighborhood of the vortex $j$, we get essentially the same state (up to an $e^{i\pi/2}$ phase) that we get when we operate with $\gamma_{j+N}$, which is an operator that creates and annihilates a particle at the neighborhood of the vortex $j+N$. Clearly, this state involves some long range correlations between the local occupation of single particle states near the cores of the vortices.

How can we move around within this subspace of degenerate ground states? Can we design some kind of a perturbation to the Hamiltonian that, if tuned properly, will take us from one ground state to another? That is a very subtle question. Generally speaking, we could think of several types of perturbations. The first is a perturbation that takes one of the particles in the super-conductor and moves it from one single-particle state to another. An example to that would be an interaction of the composite fermion with an externally applied electro-magnetic field that does not involve a flip of the spin. The second is a perturbation that scatters a composite fermion (or an electron) out of the super-conductor, for example by a spin-flip. And the third, the most interesting one, involves an adiabatic motion of the vortices. Remarkably, a perturbation of the first type would have no effect on the ground state subspace. Typically, a perturbation of this sort looks like, 
\begin{equation}
H_{pert}=\int d{\bf x}\int d{\bf x'}F({\bf x}-{\bf x'})\psi^\dagger({\bf x})\psi({\bf x'})b({\bf x},{\bf x'})
\label{ephotint}
\end{equation}
with the range of $F({\bf x}-{\bf x'})$ being very short and with $b$ being some kind of a bosonic operator (e.g., creation and annihilation of a photon or a phonon).  Most generally, the projection of $H_{pert}$ onto the ground state subspace is
\begin{equation}
H_{pert}^{gs}=\sum_{ij}h_{ij}\gamma_i\gamma_j
\label{hpertproj}
\end{equation}
When the $\psi^\dagger({\bf x})$ and $\psi({\bf x'})$ are expressed as a superposition of $\Gamma_E$ and $\Gamma_E^\dagger$ of Eq. (\ref{GammaH}), the proximity of the points ${\bf x,x'}$ to one another implies that both will have a significant overlap with the {\it same} Majorana operator $\gamma_i$. As a consequence, in Eq. (\ref{hpertproj}), as long as the vortices are well separated from one another, $h_{ij}\propto \delta_{ij}$, making Eq. (\ref{hpertproj}) nothing but a c-number (remember that $\gamma_i^2=1$ for all $i$'s). Thus, as long as it is local, the perturbation does not induce transitions from one ground state to another and does not remove the degeneracy of the ground states. 

The stability of the degeneracy of the ground state with respect to local perturbations may also be understood from the point of view of the Bogolubov-deGennes equations. Since the spectrum of Eq. (\ref{BdG}) is symmetric with respect to $E=0$, for a perturbation to affect the degeneracy of the ground states it must mix two $E=0$ solutions of the BdG equations into two solutions of non-zero energies $\pm E$. For that to happen, however, the perturbation must have matrix elements that couple the two zero energy solutions. Since those are localized within vortex cores, as long as vortices are well separated the removal of the degeneracy will be exponentially small. 

Having the super-conductor shift from one ground state to another by a perturbation that changes the parity of the number of particles in the super-conductor is easier to do. For example, consider a perturbation of the type (\ref{ephotint}) in which the spin state that is created by $\psi^\dagger$ is opposite to that annihilated by the $\psi$. Such a perturbation, when translated to the BdG operator, would involve only a single BdG operator, and thus would not assume the form (\ref{hpertproj}). Note, however, that if the spin polarized super-conductor is the ground state of the system, taking a particle out of it may involve an energy cost, and thus put the entire system in an excited state.

The most interesting way to evolve the super-conductor from one ground state to another is through braiding of vortices\cite{Ivanov01,Nayak96c,Stern04}. Look first at the case where vortex $i$ encircles vortex $j$: independent of the geometry of the trajectory taken by the vortices, when that happens  ${\rm arg}({\bf r_i}-{\bf r_j})$ changes by  $2\pi$. Hence both $\Omega_i$ and $\Omega_j$ change by $\pi$, and both $\gamma_i$ and $\gamma_j$ are multiplied by $-1$. {\it Note, these are changes in operators}. The unitary transformation that operates on the ground state when vortex $i$ encircles vortex $j$, call it $U_{ij}$, can be regarded as transforming the operators $\gamma_k$ according to 
\begin{equation}
\gamma_k\rightarrow U_{ij}^\dagger\gamma_k U_{ij}
\label{unitarytrans}
\end{equation}
The requirement that this transformation multiplies $\gamma_i$ and $\gamma_j$ by $-1$ while leaving all other $\gamma$'s unaltered is enough to fix $U_{ij}$ up to an abelian phase to be, 
\begin{equation}
U_{ij}=\gamma_i\gamma_j=\exp{\pi\gamma_i\gamma_j/2}=\gamma_i\gamma_j
\label{winding}
\end{equation}

The effect of vortex $i$ going around vortex $j$ must be identical to the effect of two interchanges of vortices $i$ and $j$, and thus it is not surprising that (up to an abelian phase again) the unitary transformation that corresponds to an interchange is the square root of $U_{ij}$ namely, $\exp{\pm\pi\gamma_i\gamma_j/4}$, where the sign is determined by the sense of the interchange.  Unitary transformations that correspond to different interchanges do not necessarily commute (in fact, two such transformations do not commute if they share one vortex), and hence the vortices satisfy a non-abelian statistics.  

So to summarize this section, the combination of a degenerate ground state subspace and a topological unitary transformation that is applied to the state of the super-conductor whenever vortices braid makes the vortices satisfy a non-abelian statistics. Remember that the super-conductor we think of is the $\nu=5/2$ state and the vortices are the quasi-particles and quasi-holes of that state. Would similar phenomena occur in two dimensional superfluid He-3 of the $p_x\pm ip_y$ pairing, or in two dimensional super-conductors of that symmetry? We will not get into a detailed discussion of this question  here, but merely comment on two important differences between the quantum Hall $\nu=5/2$ state and the latter two examples: first, as we commented earlier, two dimensional super-fluids and super-conductors have a gapless bulk mode in the form of a charge density wave. And second, their vortices do not carry electrical charge and therefore cannot be manipulated with the help of a magnetic or electric field. The Majorana intra-vortex states and the ground state degeneracy they create are both there and have measurable consequences, but the question of the statistics of the vortices is more subtle. 

\section{The geometric phases and quantum entanglement behind the non-abelian statistics}

Before we look into the way non-abelian statistics may be observed through interferometry, we will dwell a little bit on the meaning of the unitary transformations that are applied when vortices interchange, or encircle one another. These transformations were first worked out using Conformal Field Theory by Nayak and Wilczek \cite{Nayak96c}, and then using the $p$-wave super-conductor description described above by Ivanov \cite{Ivanov01} (see also \cite{Stern04,Chung07,Stone06,PhysRevLett.90.016802}). Look again on the unitary transformation that results from vortex $j$ encircling vortex $j+1$ (Eq. (\ref{winding})). Written in full detail, it is 
\begin{equation}
\left( c_{j}e^{\frac{i}{2}\Omega
_{j}}+c_{j}^{\dagger}e^{-\frac{i}{2} \Omega _{j}}\right) \left(
c_{j+1}e^{\frac{i}{2}\Omega _{j+1}}+c_{j+1}^{\dagger}e^{-\frac{
i}{2}\Omega _{j+1}}\right ),
\label{unit-trans}
\end{equation}
where the operators $c_{j}^{(\dagger)},c_{j+1}^{(\dagger)}$
annihilate (create) a composite fermion localized very close to the cores
of the $j$th and $(j+1)$th vortex, respectively. 
Eq.\
(\ref{unit-trans}) seemingly implies that the motion of the $j$th
vortex around the $(j+1)$th vortex affects the occupations of
states very close to the cores of the two vortices. This is
in contrast, however, to the derivation leading to Eq.\ (\ref
{unit-trans}), which explicitly assumes that vortices are kept far
enough from one another so that tunneling between vortex cores may
be disregarded. How can that happen \cite{Stern04}?

In fact, no tunnelling takes place. Rather, as we now explain, two
ingredients are essential for the nonabelian statistics of the
vortices. The first is the {\it quantum entanglement} of the
occupation of states near the cores of distant vortices. The second
ingredient is familiar from our discussion of abelian anyons: the {\it
  geometric phase} accumulated by a vortex traversing a closed loop.
  
As we saw, the anyonic statistics in the $
\nu=1/m$ state is related to the
geometric phase accumulated by a vortex traversing a close trajectory.
Roughly speaking, the vortex accumulates a phase of $2\pi$ per fluid
particle which it encircles.  When another vortex with its fractional
charge is introduced to the encircled area, this phase changes by a
fraction of $2\pi$, due to the fractional charge carried by the vortex.  Upon adapting the argument to interchanging of
vortices, one finds that this fraction of $2\pi$ translates into
fractional statistics.

Similarly, the Moore-Read theory of the $\nu=5/2$ state describes it also as a
superfluid, with the quasiparticles being vortices in that superfluid.
However, the ``effective bosons'' forming the superfluid are Cooper
pairs of composite fermions. Consequently, the superfluid has
excitation modes associated with the breaking of Cooper-pairs.  In the
presence of vortices, a Cooper-pair may be broken such that one or two
of its constituents are localized in the cores of vortices. For
$p-$wave superconductors, the existence of zero-energy intra-vortex
modes leads, first, to a multitude of ground states, and, second, to a
particle-hole symmetric occupation of the vortex cores in all ground
states. When represented in occupation-number basis, each of the ground states we deal with is
a superposition which has equal probability for the vortex core being
empty or occupied by one composite fermion. What distinguishes ground states from one another is the relative phases between the different components of the super-position.

When a vortex traverses a trajectory that encircles another vortex,
the phase it accumulates depends again on the number of fluid
particles that it encircles. Since a fluid particle is in this case a
Cooper pair, the occupation of a vortex core by a fermion, half a
pair, leads to an accumulation of a phase of $\pi$ relative to the
case when the core is not occupied. It is this relative phase of $\pi$ that the encircling introduces between the different components of  the wave function that might transform the system from one ground state to another.

This line of thoughts leads to the following distinction between three types of fractional quantum Hall states. The Laughlin $\nu=1/m$ fractions are condensates of one type of particle, the composite bosons formed by the attachment of $m$ flux quanta to each electron\cite{Zhang89}. The Jain states are several condensates living in parallel, each condensate being formed of a different type of composite boson. The non-abelian states are bosons made up of several electrons with flux attached to them. The most prominent example of the third type is the $\nu=5/2$ state, but this state is followed by a whole series, the Read-Rezayi series. In Section (\ref{non-abelian-wf}) we will use this line of thoughts to describe the Read-Rezayi series.

\section{Interferometry of non-abelian anyons\label{interferometers-nonabelian}}

As we described above for abelian anyons, interferometers seem to be the most direct probe of the topological interaction between anyons, since it is a geometry where winding of particles around one another is inherent in the way transport takes place. Moreover, it has recently been shown that, if realized, a Fabry-Perot interferometer of non-abelian anyons may be used as a topologically-protected qubit \cite{DasSarma05}, with an astonishingly low error rate. In this section we revisit the two geometries we discussed above with the goal of analyzing how they manifest the unique properties of the non-abelian $\nu=5/2$ state. 

\subsection{Fabry-Perot interferometer}
Again, we consider the Fabry-Perot geometry and analyze the dependence of the back-scattered current on the area of the cell between the two point contacts and on the magnetic field. The magnetic field determines the number of quasi-holes localized within the island and the area of the cell determines the number of fluid particles encircled by the interference loop. The effect of the localized quasi-particles in the non-abelian case is more subtle than in the abelian one, since the edge quasi-hole that encircles them may modify the quantum state they are in \cite{Fradkin98,SternHalperin, Bonderson06a,PhysRevLett.97.146802,PhysRevA.64.062107,Georgiev06a}. 

Let us see how this happens: suppose that $N_{qp}$ quasi-particles are localized in the cell in between the two point contacts of Fig. (\ref{interferometers}a), and that their quantum state is $|\psi_i\rangle$. Transport current flows  rightwards on the lower edge, and may be back-scattered to flow leftward on the upper edge. To lowest order, the back-scattering probability is the interference of two amplitudes, of back-scattering by the left and by the right quantum point contact. The partial wave that is back-scattered by the left point contact does not encircle the cell and thus does not modify the state $|\psi_i\rangle$. By Eq. (\ref{winding}) the partial wave that is back-scattered by the right point contact applies a unitary transformation on $|\psi_i\rangle$, whose form is, up to a phase, $\gamma_a^{N_{qp}}\prod_{j=1}^{N_{qp}}\gamma_j$, where $\gamma_j$ ($j=1..N_{qp}$) are the Majorana modes associated with the localized quasi-holes and $\gamma_a$ is the Majorana mode of the quasi-hole that is being back-scattered. Clearly, there is a difference between even and odd $N_{qp}$. When $N_{qp}$ is even, the interfering waves take the form, 
\begin{equation}
t_1|\psi_i\rangle+t_2\gamma_{cell}|\psi_i\rangle
\label{twowaves}
\end{equation}
where $\gamma_{cell}\equiv \prod_{j=1}^{n_{is}}\gamma_j$, and $t_1,t_2$ are again the amplitudes for tunnelling at the two point contacts. The interference term is then 
\begin{equation}
2{\rm Re}\left [ t_1^*t_2\langle\psi_i|\gamma_{cell}|\psi_i\rangle\right ]
\label{interferenceterm}
\end{equation}
There are two possible eigenvalues to $\gamma_{cell}$, either $\pm 1$ or $\pm i$, depending on whether $N_{qp}$ is divisible by four. Thus, as the area of the cell is varied, there are two possible interference patterns, that differ by a $\pi$ phase shift. If $|\psi_i\rangle$ is an eigenvector of $\gamma_{cell}$ the observed interference pattern would follow the corresponding eigenvalue. If that is not the case, the back-scattered current would measure the value of $\gamma_{cell}$ and one of the two interference pattern would emerge. For an even $N_{qp}$ there are $2^{N_{qp}/2}$ ground states, characterized by $N_{qp}/2$ quantum numbers, each taking one of two possible values. The Fabry-Perot interferometer measures just {\it one} of these numbers. 

As shown by Das Sarma {\it et al.} \cite{DasSarma05}, the Fabry-Perot interferometer, if brought to work for the $\nu=5/2$ state with an even $N_{qp}$, may be used as a qubit, with the two possible values of $\gamma_{cell}$ being the two states of the qubit. To flip the qubit from one value of $\gamma_{cell}$ to another, a quasi-hole should tunnel through the cell, encircling an odd numbered subset of the $N_{qp}$ localized quasi-holes. This qubit is protected from the sources of dephasing that are so harmful to other types of qubits. Perturbations such as an external RF noise, hyperfine coupling with nuclear spins etc. do not affect its behavior at all, as long as their frequencies are too low to overcome the energy gap. 

When $N_{qp}$ is odd, the unitary transformation applied by the partial wave that encircles the cell depends on the operator $\gamma_a$ of the edge quasi-particle. As current is flowing, the unitary transformations applied by that partial wave are $\gamma_a\gamma_{cell}$ followed by $\gamma_b\gamma_{cell}$ followed by $\gamma_c\gamma_{cell}$ etc., with $\gamma_a,\gamma_b,\gamma_c$ being the Majorana operators associated with the quasiparticles that flow along the edge. As these unitary transformations do not commute with one another, their expectation values will be a series of random numbers of magnitude equal or smaller to one. Consequently, no interference pattern will be observed. Rather than having phase jumps as the magnetic field is varied and $N_{qp}$ is changed, as we saw for the abelian case, the $\nu=5/2$ case is characterized by regions of magnetic field where no interference is to be seen and regions where the interference assumes one of two possible patterns. 

As the back-scattering amplitudes get bigger, higher-order reflection amplitudes should be summed up. By the argument above, for an even $N_{qp}$ all orders contribute, while for an odd $N_{qp}$ the odd harmonics, in which the interference loop winds the $N_{qp}$ quasi-particles an odd number of times, do not interfere coherently. The back reflected current then takes the form, 
\begin{equation}
I=\sum_{m=0}^\infty I_m \cos^2\left [ {mN_{qp}\frac{\pi}{2}} \right ] \cos{m\left[\phi+\frac{N_{qp}\pi}{4}+\chi_m(N_{qp})\right]}
\label{ibrna}
\end{equation}
when $N_{qp}$ is even $\chi_m(N_{qp})=\pm \frac{m\pi}{2}$ according to the eigenvalue of $\gamma_{cell}$. When $N_{qp}=2j+1$ (with $j$ an integer) then $\chi_m=j\pi$.  The calculation of the prefactors $I_m$ becomes complicated for large $m$, but is not needed for our purpose.

\begin{figure}
\includegraphics[width= 0.6\linewidth,angle=-90]{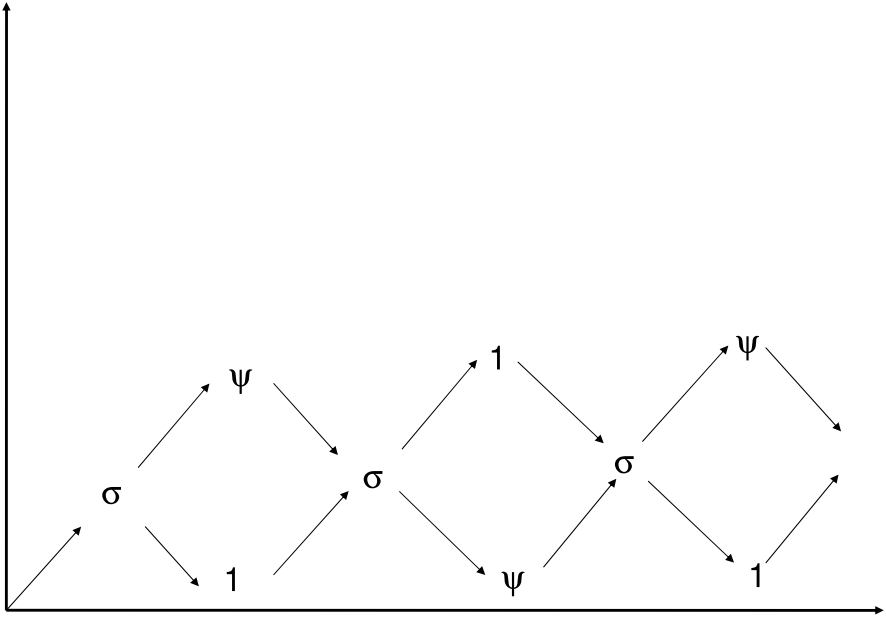}
\vspace{5mm}\caption{\label{Brattelli} A Brattelli diagram as a way to describe interference at $\nu=5/2$. The $x$-axis counts the number of quasi-particles in the interference loop. The $y$-axis describes the state of the quasi-particles. For an even $N_{qp}$ there are two possible interference patterns, mutually shifted by $\pi$, and the interferometer may be utilized as a qubit \cite{DasSarma05}. For $N_{qp}$ there is zero amplitude of interference in lowest order of tunnelling. An ideal Fabry-Perot interferometer stays on one of the nodes indefinitely. A noisy Fabry-Perot interferometer, where $N_{qp}$ fluctuates, diffuses on the Brattelli diagram. A Mach-Zehder interferometer propagates on the diagram: each tunnelling quasi-particle changes $N_{qp}$ by one. The notation $1,\psi,\sigma$ is explained in the text.}
\end{figure}

Similar to the abelian case, when the number $N_{qp}$ fluctuates in time the visibility of the Aharonov-Bohm oscillations develops a power-law relation with the average current, due to the averaging of the harmonics $m$ in (\ref{ibrna}) that do not divide by $4$. In that case the visibility increases with the increase of the average current, and the fluctuations in $N_{qp}$ result in a current noise \cite{Grosfeld06b}.

\subsection{Mach-Zehnder interferometer}

Again, as for the abelian case, the main difference between the Fabry-Perot and Mach-Zehnder interferometers is in the fact that for the former $N_{qp}$ is a parameter whose dynamics is (at least as a matter of principle) independent of the current flowing in the interferometer, while for the latter $N_{qp}$ changes by one with each tunnelling quasi-hole.  Similar to the abelian case, the time evolution of the system is a stochastic process, but now the state of the system is not characterized only by the number of quasi-holes that have tunnelled from the exterior to the interior, but also by the state of the system: when $N_{qp}$ is even there are two possible states, and a different transition rate corresponds to each one of them. This state of affairs can be diagrammatically expressed by means of Brattelli diagrams, see Fig. (\ref{Brattelli}). The $x$-axis of the diagram is $N_{qp}$. The $y$-axis is the state of the cell, i.e., the interference pattern it shows (this is a rather careless definition of the $y$-axis. A more elaborate one will be given in the Section (\ref{non-abelian-wf})). For reasons that have to do with conformal field theory, and will be explained below, the state of odd $N_{qp}$ is commonly called $\sigma$, while the two states of even $N_{qp}$ are called $1$ and $\psi$ (or sometimes $1$ and $\epsilon$). When $N_{qp}$ is odd, the system is in the state $\psi$ and may go to either $1$ or $\psi$. The transition rates to these states differ from one another since the relative phase of the two interfering waves depends on the final state. When $N_{qp}$ is even, there are two possible states the system may be in ($1$ and $\psi$), and one state to which it may go. Again, two different scattering rates. It turns out that altogether there are four transition rates, which depend on the number $N_{qp}{\rm mod} 4$ also due to the abelian phase accumulated around the interference loop. The detailed calculation of the various transition rates is easiest to carry out using the techniques that will be explained below, in Sec. (\ref{non-abelian-wf}) \cite{Feldman06}.

The evolution of the state of the system is determined by a set of equations similar to (\ref{rateequation}), with results that are similar in spirit, but different in important details: the current is periodic in the flux through the interferometer, with a period of $\Phi_0$. The visibility scales as a power law of the average current, with the power being unique to the $\nu=5/2$ state. And the effective charge (Fano factor) measured in a shot noise measurement depends on the flux as well\cite{Feldman07}. 

The range of effective charges (Fano factors) to be seen in various values of the flux is a signature of the $\nu=5/2$ state. The dependence of the effective charge on the flux is, as we saw for the abelian case, a consequence of the existence of several scattering rates. Similar to the abelian case, two cases are rather easy to analyze. When all transition rates are the same, the effective charge is that of the quasi-hole, namely $e/4$. When one transition rate is significantly smaller than all others, again the effective charge will be larger, since the system will be stuck for long times in the state from which it is hard to move on, and then move with a burst of steps until the next time it gets stuck. The structure of the Brattelli diagram allows the system to "bypass" the state from which the transition rate is small if that rate happens on a transition out of $\sigma$, and thus does not limit the number of steps in the "burst".  This structure allows for an effective charge that is larger than the electronic charge. In fact, in the case of the $\nu=5/2$ state the largest Fano factor should be higher than three electronic charges. 

Another scheme that uses current noise for the observation of signs of non-abelian statistics was suggested in \cite{PhysRevB.73.155335}.

\subsection{Resonant tunnelling through a quantum dot}

When the two quantum point contacts in a Fabry-Perot interferometer (Fig. (\ref{interferometers}a) are tuned to strong back-scattering, the cell between the point contacts becomes effectively a quantum dot, where the number of electrons is quantized. Equivalently, in this regime the conductance of the interferometer may be viewed as determined by multiple windings of the cell, induced by multiple reflections by the point contacts. Rather than having a sinusoidal dependence of the back-scattered current on the area of the dot, we then expect to have most of the current back-scattered by the point contacts, except in those values of the area where resonant tunnelling takes place and all the current is transmitted through the dot. Those values correspond to area where the ground state energy of the dot with $N$ and with $N+1$ electrons is degenerate. 

In view of this condition, both for integer quantum Hall states and for abelian fractional quantum Hall states we expect the difference in area between two consecutive resonances to be $1/n_0$, where $n_0$ is the electronic density in the dot. This prediction is little more than the observation that when the dot has several thousand electrons, the physics of the degeneracy point between the different nearby values of $N$ is identical. One needs to increase the area of the dot by $1/n_0$ in order for it to accommodate another electron.

The non-abelian $\nu=5/2$ state is different, since in that state electrons are paired. Thus, we may wonder whether the periodicity of the peaks on the area axis would be the area needed for accomodating another pair, or that needed for accommodating another electron, $2/n_0$ or $1/n_0$ (here $n_0$ is the density of electrons in the half filled uppermost Landau level). A similar question arises, of course, for super-conductors, where the peaks are scanned on the axis of a gate voltage rather than area. For super-conductors, the spacing between peaks indeed alternate, depending on whether the added electron has an electron in the dot to pair with, or whether it remains unpaired. The spacings in the two cases differ by the energy gap of the super-conductor. 

In the $\nu=5/2$ case, the answer depends on $N_{qp}$\cite{SternHalperin}. If $N_{qp}$ is odd, there is a zero energy mode on the edge of the dot. This is, in the language of a super-conductor, a mid-gap state that is available for the tunnelling electron, since it lies at the edge. Thus, in this case the spacing between transmission peaks will not alternate, and the periodicity in area will be $1/n_0$. In contrast, this zero energy state is absent from the edge when $N_{qp}$ is even. As long as there is no edge-bulk coupling  the tunnelling electron cannot make use of the zero energy states in the bulk, and the resonant peaks alternate: an odd electron must reside on the edge, occupying the lowest energy edge mode, while an even electron forms a Cooper-pair and is absorbed by the bulk. Thus, when $N_{qp}$ is even, the periodicity in area becomes $2/n_0$. Even when the measurement is of a transmission through a quantum dot, where the number of {\it electrons} is quantized, the combined dependence on area and magnetic field carries a signature of the non-abelian nature of the {\it quasi-holes}.

\section{Other non-abelian states\label{non-abelian-wf}}


The $\nu=5/2$ state is the simplest example of a non-abelian quantum Hall state. There are several ways to approach the others. A possible starting point, which we choose here, is that of trial wave functions. Despite the obvious difficulty in constructing a good approximation for an eigenstate wave function of a system of about $10^8$ interacting electrons, trial wave functions  led to enormous progress in the study of the fractional quantum Hall effect, as we will review. Other starting points, originating from the non-abelian Chern-Simons theories and from quantum groups, are covered extensively in \cite{RMP} and \cite{Slingerland01}. 

Our journey towards non-abelian quantum Hall states on the trial wave function route starts early, with the construction of the ground state wave function for a system of non-interacting electrons at $\nu=1$. These electrons would fill up the lowest Landau level, with all spins polarized, and would form a many-body wave function that is just the Slater determinant of all single particle lowest Landau level states. In the symmetric gauge the single particle states are characterized by their angular momentum $m$. When the $i$'th electron occupies the $m$'th state its single particle wave function is $z_i^m\exp{-\frac{|z_i|^2}{4l_H^2}}$ and the many-body Slater determinant becomes, 
\begin{eqnarray}
\psi_{\nu=1}\left (\{z_i\}\right) &= &{\rm Det}z_i^m\exp{-|z_i|^2/4l_H^2}\\ & = & \prod_{i< j}(z_i-z_j)\prod_i\exp{-|z_i|^2/4l_H^2}
\label{nueq1wf}
\end{eqnarray}
where in the first equality $i,m$ are the indices of the matrix.

This wave function has some properties built into it by construction:  it is purely within the lowest Landau level, it describes a $\nu=1$ droplet, and it is fermionic (anti-symmetric to the exchange of two of its coordinates). The wave function is exact for non-interacting electrons, and is expected to be a good approximation when the inter-electron interaction energy is much smaller than the cyclotron energy. 

The next stop on our journey are the celebrated Laughlin wave functions. Let us generalize Eq. (\ref{nueq1wf}) by defining 
\begin{equation}
\psi_{\alpha}\left (\{z_i\}\right) = \prod_{i< j}(z_i-z_j)^\alpha\prod_i\exp{-|z_i|^2/4l_H^2}
\label{nualphawf}
\end{equation}
When $\alpha$ is an odd integer, this is a Laughlin wave function, but we will think here of $\alpha$ as a parameter to be varied. Generally, the value of $\alpha$ determines the filling factor. This may be understood either by observing that the probability distribution for finding a single electron at a distance $|z|$ very far from the origin scales like $|z|^{2\alpha(N-1)}e^{-|z|^2/2l_H^2}$, with $N$ being the number of electrons, thus making the effective area of the droplet scale like $\alpha N$ and the filling factor scale as $1/\alpha$, or by means of the plasma analogy. Writing $|\psi\left (\{z_i\}\right )|^2\equiv \exp{-{\cal H}}$, with 
\begin{equation}
{\cal H}=2\alpha\sum_{i<j}\log{(z_i-z_j)}-\sum_i|z_i|^2/2l_H^2,
\label{plasma}
\end{equation}
the function $\cal H$ can be regarded as the classical Hamiltonian of a fictitious two dimensional plasma, and the electronic probability distribution as the partition function corresponding to that plasma. In this two-dimensional plasma the fictitious particles, whose coordinates are $z_i$, mutually interact through a $\alpha\log(z)$ interaction, and interact with a uniform background density (that in two dimensions creates a quadratic potential). The condition of charge neutrality then fixes the density to be $1/2\pi \alpha l_H^2$ and the filling factor to be $1/\alpha$.

If we combine together: {\it (a).} the derivation of  Eq. (\ref{nueq1wf}) as the exact ground state for non-interacting electrons at $\nu=1$, {\it (b)}. the observations made above regarding the wave function (\ref{nualphawf}), and {\it (c)} the requirement that for filling factors smaller than one, in the limit of a very large cyclotron energy, the ground state wave function should be constructed of states of the lowest Landau level only; And if we add to this list a lot of Laughlin--inspired  hindsight, then it becomes almost natural to use the wave function (\ref{nualphawf}) with $\alpha=3,5$ as trial wave functions for the $\nu=1/3,1/5$ ground states.  As long as $\alpha$ is an odd number, the wave function
(\ref{nualphawf})
is of fermionic symmetry and describes - as seen above using the plasma analogy - a droplet of filling factor $\nu=1/\alpha$. Its exceptional success in describing the actual $\nu=\frac{1}{3},\frac{1}{5}$ states may be ascribed to the clever use it makes in its zeroes. Other than the exponential factors, this wave function is a polynomial in the $z_i$'s. The largest power it has for each $z_i$ is determined by the filling factor to be of order of $\alpha N$. Thus, this is also the number of zeroes for each $z_i$'s.  Surely, there must be a zero whenever two electrons are in the same position, due to the Pauli principle of Fermions. That forces the position of $N-1$ zeroes per each coordinate, but does not force the position of the other $(\alpha-1)(N-1)$ zeroes per coordinate. The Laughlin wave function devotes all its zeroes to keeping electrons away from one another, and thus minimizes their probability for being close to one another. As the distance $\delta z$ between two electrons gets smaller and smaller, the probability to find them at that distance decreases as $|\delta z|^{2\alpha}$, faster than the $|\delta z|^2$ forced by the Pauli principle. 

What happens, however, when $\nu=1/\alpha$ and $\alpha$ is not an odd integer? There are two approaches that have been taken for constructing wave functions for this case: Jain's composite fermions approach and Read-Rezayi's clustering approach. The composite fermions approach leads to a series of abelian quantum Hall states, while the clustering approach leads to a series of non-abelian states. For those filling fractions for which both abelian and non-abelian trial wave functions may be constructed, the details of electron-electron interaction determine whether one of the two, and which one, is indeed the phase of lowest energy.

Jain's approach, which has been the impetus for the development of the composite fermions field theory described above in Sec. (\ref{cftheory}), starts from the following decomposition of $\psi_\alpha$ for any odd $\alpha$ and $\nu=1/\alpha$:
\begin{eqnarray}
\psi_\alpha=\prod_{i<j}(z_i-z_j)^{\alpha-1}\prod_i\exp{-2\pi n|z_i|^2(\alpha-1)}\nonumber \\ \times\quad \prod_{i<j}(z_i-z_j)\exp{-2\pi n|z_i|^2}
\label{jain}
\end{eqnarray}
Given that $\l_H^{-2}=2\pi n\alpha$, where $n$ is the electronic density, this breaks the wave function into two factors. The first one, to the left of the $\times$ sign, attaches an even number $(\alpha-1)$ of flux quanta to each electron, transforming it into a composite fermion. The filling factor of the composite fermions is $1$ and the second factor describes their occupation of a single Landau level of a reduced magnetic field. Based on this observation, Jain constructed the wave function for $p$ filled Landau levels of composite fermions, corresponding to a filling factor of $\frac{p}{(\alpha-1)p+1}$ by replacing the single Landau level wave function, the second line  in (\ref{jain}), by the wave function for $p$ filled Landau levels, and then projecting the resulting wave function into the lowest Landau level. The resulting wave functions are identical in their topological properties to those obtained from the field theoretical approach for composite fermions. Specifically, the quasi-particles and quasi-holes follow abelian statistics.

Read-Rezayi's clustering approach, which is our focus in this Section, starts by considering the wave function (\ref{nualphawf}) for a general value of $\alpha$.  This wave function describes a droplet of the correct filling factor, but - lamentably - for an even $\alpha$ it does not have the proper fermionic symmetry, and for a non-integer $\alpha$ it is not even single-valued. As it stands, then, Eq. (\ref{nualphawf}) is not an acceptable trial wave function for electrons in a general filling factor $\nu=1/\alpha$. 

Can this be fixed? What does it take to fix it? We look for a trial wave function of the form, 
\begin{equation}
\Psi\left (\{z_i\}\right)=\psi_\alpha\left (\{z_i\}\right){\cal F}\left (\{z_i\}\right)
\label{cftprecursor}
\end{equation}
where the function $\cal F$ will fix all the harm done by a non-odd value of $\alpha$, without varying the filling factor. For the relatively easy case of an even $\alpha$, we need to find a function ${\cal F}\left (\{Z_i\}\right)$ that is odd with respect to interchanging any of its two arguments, yet does not involve powers of $z_i$ that are of order $N$. For the harder case of a non-integer $\alpha$ the function $\cal F$ should guarantee the single-valuedness and fermionic symmetry of  $\Psi$. To do that, it should scale like $(z_i-z_j)^{-\beta}$ when $z_i\rightarrow z_j$, with $\alpha-\beta$ an odd positive integer. Yet, again, it should not involve high powers of each of the coordinates. The Read-Rezayi approach searches for such a function.

Following the rationale of flux attachment as a means to map problems of electrons in a magnetic field onto composite particles at a different magnetic field, we may view, at least roughly, the wave function $\psi_\alpha$ (Eq. (\ref{nualphawf})) as a flux attachment transformation that attaches $\alpha$ flux quanta to each electron. For a filling fraction $\nu=1/\alpha$ that makes the composite particles experience zero magnetic field on average. Thus, in Eq. (\ref{cftprecursor}) $\Psi$ is the electronic wave function, $\psi_\alpha$ is the flux attachment transformation, and the function ${\cal F}$ is the wave function of the composite particles, that carry one electron charge and $\alpha$ flux quanta, and are subjected to zero average magnetic field. For an odd $\alpha$ the electrons are turned into bosons at zero average magnetic field, whose wave function is just a constant. This viewpoint of the Laughlin fractions (as well as the $\nu=1$ completely filled lowest Landau level), commonly known as composite boson theory \cite{Zhang89}, views the Laughlin fractions as Bose condensates, characterized - as all condensates - by the dissipationless flow of currents. For an even $\alpha$ the electrons are transformed into composite fermions. The function $\cal F$ is then supposed to be the function of spin polarized fermions at zero field, and as we saw above, they may form a $p$-wave super-conductor. The $p$-wave super-conductor is again a condensate. This time the condensate is formed by Cooper-pairs, a cluster of two fermions that, effectively, group together to form a boson. 

For a non-integer $\alpha$ the flux attachment turns electrons into anyons, and the function $\cal F$ is to be the wave function of anyons at zero magnetic field. For these anyons to form condensates, where current flows with no dissipation, they have to form effective bosons, made of clusters of an integer number $k$ of anyons. If the phase accumulated when one anyon encircles another is $4\pi/k$, the phase accumulated by an anyon encircling a cluster of $k$ anyons will be $4\pi$, and that accumulated by two clusters encircling one another will be a multiple of $4\pi$. The cluster will then be an effective boson. Thus, for forming a condensate of this sort, the flux attachment transformation should attach $(1+\frac{2}{k})$ flux quanta to each electron, which would make the electronic filling factor $k/(k+2)$. 

When the Laughlin fractions $\nu=1/m$ are described as condensates of bosons, the Laughlin quasi-particles are the quantized vortices in this condensate. To form a quantized vortex at the point $\bf R$ in a superfluid of bosons we need to award each boson an angular momentum of $\hbar$ relative to the point $\bf R$. When the boson is made of two fermions, the relative angular momentum per fermion is $\hbar/2$. When the fermion goes around the vortex, then, it accumulates a phase of $\pi$. In analogy, when the boson is made of $k$ anyons, each anyon gets, on average, an angular momentum of $\hbar/k$ and an anyon that goes around a vortex accumulates a phase of $2\pi/k$. 

The arguments of the last two paragraphs tell us that there should be a description of the $\nu=k/(k+2)$ fractional quantum Hall state as a condensate of effective bosons made of clusters of $k$ anyons. Within this description an anyon that goes around another would accumulate a phase of $4\pi/k$ while an anyon that goes around a vortex in the condensate would accumulate a phase of $2\pi/k$. Our considerations do not tell us, however, the information we need about a collection of vortices: what is the phase accumulated when vortices go around one another, what are the degenerate ground states, if any, that many vortices form when they are distant from one another, and how these degenerate ground states lead to non-abelian statistics, if at all. Indeed, for the case of $\nu=5/2$, described in details earlier, we needed the machinery of super-conductivity, particularly the Bogolubov-deGennes equations for a $p$-wave super-conductor, to give us that information. The tool that will help us to analyze the clustering of anyons into effective bosons and the states formed by vortices in the condensate of these bosons will be parafermionic Conformal Field Theories. As we now review, these theories provide a tool for constructing the function $\cal F$ both in the absence and in the presence of quasi-particles in the ground state.

We start with an introduction to the Conformal Field Theories we will use (Our introduction is very short and very incomplete. Useful references for a more comprehensive study of the subject are \cite{ZF,Gepner:1987sm,Gepner86,Gepner87,Senechal97,Cappelli99,Cappelli01,Verlinde88}. Following that, we show how the Ising CFT reproduces our analysis of the $\nu=5/2$ state, and finally we show how a certain set of parafermionic CFTs generate the Read-Rezayi states for $\nu=2+k/(k+2)$. When analyzing conformal field theories, general considerations of symmetries and internal consistency go a long way towards the calculation of certain properties of correlation functions. Here these correlation functions will construct the trial wave functions for the ground state and the quasi-particles. What is needed for these calculations are three inputs. The first is the list of fields in the theory. Some of these fields we may anticipate: the field of the anyon, which we will denote by $\psi_1$ (this field is called a parafermion in the CFT context), the field of the vortex, which we will denote by $\sigma_1$ and the field of the boson that is formed by $k$ anyons. As we will see below, at $\nu=5/2$ these are the only three fields, while in the more complicated CFTs there are more fields. The second input we need is the conformal dimension of each field. The third is the list of operator product expansion rules. The operator product expansion rules tell us that when two fields $\Phi_\alpha(z_1)$ and $\Phi_\beta(z_2)$ appear together in a correlation function, and when the coordinates $z_1$ and $z_2$ approach one another, the following substitution may be carried out:
\begin{equation}
\Phi_\alpha(z_1)\Phi_\beta(z_2)\rightarrow \sum_\gamma C_{\alpha\beta\gamma}(z_1-z_2)^{-\delta}\Phi_\gamma
\label{ope}
\end{equation}
where $C_{\alpha\beta\gamma}$ are constants and $\delta=h_\gamma-h_\alpha-h_\beta$ with the $h$'s being the conformal dimensions of the respective fields. The fields $\Phi_\alpha$ and $\Phi_\beta$ are then said to be "fused" to the field(s) $\Phi_\gamma$. These rules for fusion are very important for us, in several contexts. First, we will use them to "create" clusters of anyons. Indeed, we will see that when $k$ $\psi_1$'s are fused together, the resulting field is the identity. The identity field is an "invisible" one: when it is fused with any field, $\delta=0$, and the right hand side of (\ref{ope}) does not have the $(z_1-z_2)$ factor. Second, we will use the fusion rules to figure out the phase accumulated when one field encircles another. This phase is simply $2\pi\delta$, with the $\delta$ that correspond to the fusion of the two fields. And third, we will use the rules to figure out the structure of the ground state subspace created when several quasi-particles (vortices) are fused together.

So here is how it will go \cite{MR,Read99}: we will have a description of the electron as a product of two conformal fields, 
\begin{equation}
\psi_e(z)=e^{i\sqrt{\frac{k+2}{k}}\phi(z)}\psi_1(z)
\label{electrondecomposition}
\end{equation}
In the first piece, $e^{i\sqrt{\frac{k+2}{k}}\phi(z)}$, commonly called a "vertex operator", the field $\phi(z)$ is a free chiral boson. 
The ground state wave function will be a correlator of many electron fields 
\begin{equation}
\Psi(\{z_i\})=\langle\psi_e(z_1)...\psi_e(z_n)\rangle.
\label{cftwf}
\end{equation}
 The correlation functions of the bosonic  vertex operators produce Eq. (\ref{nualphawf}). The correlation function of the $\psi_1(z)$'s will produce the function $\cal F$. 
There are two important properties that will make it happen. Firstly, the fusion of a finite number $k$ of $\psi_{1}$'s will produce the identity field. That will guarantee that the power of each $z_i$ in the function $\cal F$ will scale like $k$, and not like $N$. Secondly, the conformal dimension of $\psi_{1}$ will guarantee that the entire correlator will create a fermionic wave function. Generally, the operator product expansion for the bosonic fields is, 
\begin{equation}
e^{i\alpha\phi(z)}\times e^{i\beta\phi(z')}\rightarrow e^{i(\alpha+\beta)\phi(z)}(z-z')^{\alpha\beta}
\label{bosonsfusion}
\end{equation}
Electrons must accumulate a phase of $2\pi$ when they encircle one another. They will accumulate a phase of $2\pi (k+2)/k$ from the bosonic factors, and will therefore need to accumulate $-4\pi/k$ from the parafermionic fields. This will make the combined wave function single-valued and anti-symmetric with respect to interchanging the coordinates of two electrons.

To generate vortices, we will introduce terms of $\sigma_1 e^{i\beta\phi}$ into the correlators. The value of $\beta$ will be chosen as the minimal value for which the correlators will remain single valued with respect to the electronic coordinates. As we will see, the fusion rule of two $\sigma_1$'s will be fundamentally different from the fusion rule of two $\psi_1$'s. While two $\psi_1$'s fuse to a single fusion product, two $\sigma_1$'s will have two fusion channels to go into. This multiplicity of possible outcomes will be the source of the ground state degeneracy and the non-abelian statistics. 

This description looks complicated at first reading, and - to be honest - also at later readings. It will hopefully get simpler after we  see how it works for the $\nu=5/2$ state, and then how it works for the more complicated Read-Rezayi states.

\subsection{The CFT way of analyzing the $\nu=5/2$ state}

The simplest parafermionic theory to be mentioned here is known as the $Z_2$ (or Ising) theory. It is the CFT method of dealing with the $\nu=5/2$ state. Let us see how it works: the theory has three fields:  $\psi_1$, $\sigma_1$ and the identity field. In view of that, we will omit the subscript from $\psi$ and $\sigma$. The identity field corresponds to the Cooper pair. The $\psi$ is a fermion. The $\sigma$ corresponds to a vortex. The conformal dimensions of these fields are $0,\  1/2$ and $1/16$ respectively. What makes the $Z_2$ theory simple is the fact that the action for the $\psi$ is known to be that of a free chiral Majorana fermion\cite{Senechal97}. As explained above, the electron and quasi-particle fields are composed of the parafermionic fields together with the bosonic factors.  

There are three rules for operator product expansion. For creating the wave function with no vortices we need only the following:
\begin{equation}
\psi(z)\psi(z')\rightarrow(z-z')^{-1}
\label{opepsi}
\end{equation}
In the language of super-conductivity, this says that two composite fermions may form a Cooper-pair whose wave function is $(z-z')^{-1}$. 

Let us put:
\begin{equation}
\Psi(\{z_i\})=\langle \prod_{i=1}^N e^{i\sqrt{2}\phi(z_i)}\psi(z_i) e^{-iN\sqrt{2}\phi(z_\infty)}\rangle
\label{pfaffian}
\end{equation}
with $z_\infty$ an arbitrary point at infinity, and the the last factor guarantees that the correlator of the free boson factor does not vanish. 
The free action for both $\psi$ and the boson $\phi$ enables the use of Wick's theorem, and the rule (\ref{opepsi}) allows for the explicit calculation of this correlator. The result is
\begin{eqnarray}
\psi_a(\{z_i\})=&\prod_{i<j}(z_i-z_j)^2\prod_i e^{-\frac{|z_i|^2}{4l_H^2}}\nonumber \\ &\times {\cal A}\frac{1}{z_1-z_2}\frac{1}{z_3-z_4}..\frac{1}{z_{N-1}-z_{N}}
\label{pfaffianexplicit}
\end{eqnarray}
where $\cal A$ is the anti-symmetrizer.

The part $\cal F$ of the wave function we obtain through these correlators, which is the Pfaffian of the matrix $M_{ij}=1/(z_i-z_j)$, is, in fact, the wave function for a $p_x+ip_y$ super-conductor with no vortices. Now let us look at the role of the $\sigma$'s. The operator product expansion for $\sigma$ with $\psi$ is 
\begin{equation}
\sigma(z)\psi(z')\rightarrow \sigma(z)(z-z')^{-1/2}
\label{sigmapsiope}
\end{equation}
Thus, $\sigma$ is an object that gives the fermion $\psi$ a minus sign when the fermion goes around it. Sounds like a vortex in the super-conductor, and indeed it is. The quasi-hole would be $\sigma(z)e^{i\beta\phi(z)}$, and $\beta$ is fixed in such a way that when the electron goes around a quasi-hole (or a quasi-particle) it accumulates a phase of $2\pi$ such that the wave function remains single valued. For that to happen, Eqs. (\ref{bosonsfusion}) and (\ref{sigmapsiope}) tell us that we need $\beta=\pm 1/2\sqrt{2}$, a value that fixes the charge of the quasi-hole/particle to be a quarter of the electron charge. 

When two $\sigma$'s fuse, the $Z_2$ theory has,  
\begin{equation}
\sigma(z)\sigma(z')\rightarrow (z-z')^{-1/8}\left [1+\psi(z)(z-z')^{1/2}\right ]
\label{sigmaope}
\end{equation}
Two $\sigma$'s may then be fused either to $1$ or to $\psi$, which is the CFT way of saying that two vortices in a $p_x+ip_y$ super-conductor introduce a two-fold degeneracy of the ground state, and that the two degenerate states differ by the parity of the total number of fermions. 

The only remaining piece of information about operator product expansion that we need is that the $1$ field fuses trivially with the $\psi$ and the $\sigma$, leaving them unchanged, and with no factors of $(z-z')$. 

Finally for the $Z_2$ theory, how do we understand the interferometry in the language of the CFT? Let us look at the Fabry-Perot geometry. The encircling of the bulk quasi-particles by the interference loop amounts to a trajectory in which an edge quasi-particle $\sigma(z)e^{i\phi(z)/2\sqrt{2}}$ goes around the bulk, whose $N_{qp}$ quasi-particles fuse to $e^{iN_{qp}\phi(z')/2\sqrt{2}}\xi(z')$ where $\xi(z')$ is $\sigma(z')$ when $N_{qp}$ is odd, and may be either the identity or $\psi(z')$ when $N_{qp}$ is even. 

Now, the phase accumulated when an edge quasi-particle goes around the bulk one can be obtained from their fusion rules. First, the bosonic fields yield a factor of $(z-z')^{N_{qp}/8}$. Second,  by (\ref{sigmaope}) and (\ref{sigmapsiope}), the fusion of $\sigma$ with $\xi$ gives no extra phase when the bulk fuses to the identity, and an extra phase of $\pi$ when it fuses to $\psi$. Thus, when the bulk quasi-particles fuse to the identity the edge quasi-particles accumulates only the abelian phase. This phase is just that of a vortex of a super-conductor going around $1/8$ of a Cooper-pair, namely a quarter of a charge. When the bulk quasi-particles fuse to $\psi$ an extra phase of $\pi$ is added to the phase accumulated in the previous case. Furthermore, when the bulk fuses to $\sigma$, the fusion of the edge $\sigma$ with the bulk $\sigma$ gives, according to (\ref{sigmaope}) a sum of two interference patterns of equal magnitude, with a mutual phase shift of $\pi$. Obviously, these two terms cancel one another and no interference is to be seen, exactly as predicted by the $p_x+ip_y$ superconductor picture. 

Before we turn to discuss the other non-abelian states, for which we do not have but the CFT description, let us summarize what use we made of CFT for understanding the $\nu=5/2$ state, since similar steps will be useful for the understanding of the other non-abelian states: 

\begin{enumerate}
\item We carried out a Chern-Simons transformation that is not necessarily single valued, aimed at cancelling the external magnetic field within mean field approximation. This transformation effectively maps the problem of electrons in a magnetic field onto a problem of composite particles at zero magnetic field. For $\nu=5/2$ those particles were fermions. For other filling fractions, they will be anyons. 
 
\item As the wave function $\cal F$ for the composite particles we looked for a function that will have the needed power laws when any two particles approach one another, but will be rather low in the degree of its polynomials. We then introduced the parafermionic CFT's, that happen to have exactly the needed properties.

\item We studied the $Z_2$ CFT, and used it to identify the field for the electron, the field for the quasi-hole, the degeneracy of the ground state subspace, and the phases accumulated when one type of particle goes around another.

\end{enumerate}

Note that this line of thought cannot but offer the possibility of a non-abelian phase at a particular filling fraction and analyze its properties. It cannot judge if, and at what conditions, this phase is energetically favorable to competing phases.  This question may be answered either through experiments or through numerical exact diagonalization studies. 

\subsection{The Read-Rezayi series}

Let us now apply these ideas to other filling fractions, and obtain the Read-Rezayi series. Consider a filling fraction of $k/(k+2)$. The cases $k=1,2$ are relatively easy, and were discussed at length above (True, we discussed the $\nu=5/2$ as the $k=2$, and not $\nu=1/2$. It turns out that details of the electron-electron interaction in the $N=1$ Landau level are more favorable for composite fermion pairing than those of the lowest, $N=0$, Landau level. That is apluasibly the case for the higher values of $k$ \cite{Rezayi06}). Now we look at higher integers. The Chern-Simons transformation (\ref{nualphawf}) is now multiply valued, and we resort again to the CFTs to fix this problem. Thus, we are looking for a CFT whose correlators for one of its fields scales like $(z-z')^{-2/k}$. Luckily, as we will now see, the parafermionic CFTs do just that. 

Similar to the $\nu=5/2$ case, we will combine two CFTs, a free chiral boson $\phi(z)$ and a "parafermionic" one. The parafermionic CFT will have more than the three fields we had in the $Z_2$ case. This should come as no surprise: it takes the fusion of $k$ anyons to get to the identity, so we need to have $k$ fields $\psi_j$ ($j=1..k$) where $\psi_j$ is just $j$ anyons fused together, and $\psi_k$ is the identity. Since a vortex carries a flux of $1/k$, we would also require that $k$ vortices are allowed  to fuse to the identity (but are not forced to do so. Remember that vortices have several fusion channels). That requires $k$  fields $\sigma_j$ which would be $j$ vortices fused together (the field $\sigma_k=\psi_k$ would be the identity field). But even that is not all: let us call $\sigma_{k-1}$ the anti-vortex. If the vortex and anti-vortex do not necessarily fuse to the identity, then a collection of $j$ vortices and $l$ anti-vortices does not necessarily fuse to $\sigma_{j-l}$, but may fuse to another state. That state will accumulate the same phase when encircling a parafermion, but a different phase when encircling a vortex. 

Altogether, a more elaborate analysis along these lines leads to the following picture: it turns out to be convenient to label the fields by two quantum numbers, $\Phi^l_m$. The integer $m$ is called "the topological charge". Fusion processes conserve the topological charge modulo $2k$, and it is the topological charge that determines the phase accumulated when a parafermion encircle the field $\Phi^l_m$. The integer index $l$ satisfies $l\in \left\{ 0,1,\ldots ,k\right\}$, and the fields $\Phi_m^l$ satisfy $\Phi _{m
}^{l }=\Phi _{m +2k}^{l}=\Phi _{m -k}^{k-l }$ and $l+m \equiv
0\left( {\rm mod}2\right) $. These fields include all the fields anticipated in the paragraphs above:  $\psi_1\equiv\Phi^0_{2}$ is the parafermion, which will play the role of $\psi$ in the $Z_2$ case; The fields $\psi_j\equiv\Phi^0_{2j}$ are collections of $j$ parafermions; The field
$\sigma_1\equiv \Phi^1_1$ is the vortex (called also the spin field); The field $\Phi_0^2$ is the field that may be obtained (side by side with the identity) when a vortex and and an anti-vortex fuse together; And finally $\Phi_k^k=\Phi_0^0$ which is the identity. 

The conformal dimension of the field $\Phi _{m}^{l}$ is given by  (for $-l\le m\le l$)
\begin{equation}
h_{m}^{l}=\frac{l(l+2)}{4(k+2)}-\frac{m^2}{4k}.
\label{conformaldimensions}
\end{equation}
 The operator product
expansion (OPE), is given by $\Phi_i(w)\times\Phi_j(z)=\sum_k
C_{ijk}(z-w)^{h_k-h_i-h_j}\Phi_k$, where the fields appearing on
the right hand side are determined by the equation\cite{ZF,Gepner86}: 
\begin{equation}\label{eq:fusion}
\Phi _{m _{\alpha}}^{l _{\alpha}}\times \Phi _{m _{\beta}}^{l
_{\beta}}\rightarrow\sum\limits_{l =\left| l_{\alpha}-l _{\beta}\right|
}^{\min \left\{l _{\alpha}+l_{\beta},2k-l _{\alpha}-l
_{\beta}\right\} }\Phi _{m _{\alpha}+m _{\beta}}^{l },
\end{equation}

What does this tell us? First, that 
\begin{equation}
\psi_1(z)\psi_1(z')\rightarrow\psi_2(z)(z-z')^{-2/k}
\label{anyonsfusion}
\end{equation}
 as we need. Second, that when we fuse $k$ parafermions $\psi_1$ we get the unity. Thus, the state we will construct this way will be a condensate of clusters, each cluster having $k$ parafermions. 

As we said above, dissipationless transport is a property of a bosonic system at zero magnetic field. At $\nu=1/m$ the flux attachment term (\ref{nualphawf}) turned the electrons into composite bosons. At $\nu=1/2$ the attachment of flux turned electrons into composite fermions and pairs of composite fermions acted like bosons. At $k=3..$ the electrons were turned into anyons, and the grouping of $k$ anyons creates the bosons that condense.

The identification of the electron creation operator is done as for the $\nu=5/2$ case. It is $\psi_1
e^{i\frac{\phi}{\sqrt{k}}\sqrt{k+2}}$. Similarly, the quasi-particle
creation operator is $\sigma_1
e^{i\frac{\varphi}{\sqrt{k}}\frac{1}{\sqrt{k+2}}}$. Their charge is 
$\frac{e}{k+2}$ and they carry an average angular momentum of $\hbar/k$ per particle. 

In the absence of quasi-particles this scheme creates a single ground state, which is the correlation function of many electron operators. For each $z_i$, the polynomial part of the wave function has a maximal power of about $N(k+2)/k$, and hence about that number of zeroes (the word "about" stands for the neglect of numbers of order one, and for the neglect of the question of whether $N$ divides by $k$). Out of these zeroes for, say, $z_1$, $N-1$ zeroes coincide with the positions of the other electrons, as required by the Pauli principle. The other $2/k$ zeroes are not attached to other electrons. Rather, it can be shown that if the electronic wave function $\Psi$ in Eq. (\ref{cftprecursor}) is written as $\Psi=\prod_{i<j}(z_i-z_j)\chi(\{z_i\})$ (separating out the zeroes that are dictated by the Pauli principle) then the function $\chi$ vanishes quadratically when $k+1$ of the coordinates approach one another \cite{Read99,Cappelli99, Cappelli01} . Thus, when $k$ of the coordinates are close together around the point $z_0$, we expect the electronic wave function $\Psi$ to vanish like $(z_1-z_0)^2$, attaching two zeroes - figuratively speaking - to $k$ electrons. 

\begin{figure}
\includegraphics[width= 0.6\linewidth,angle=-90]{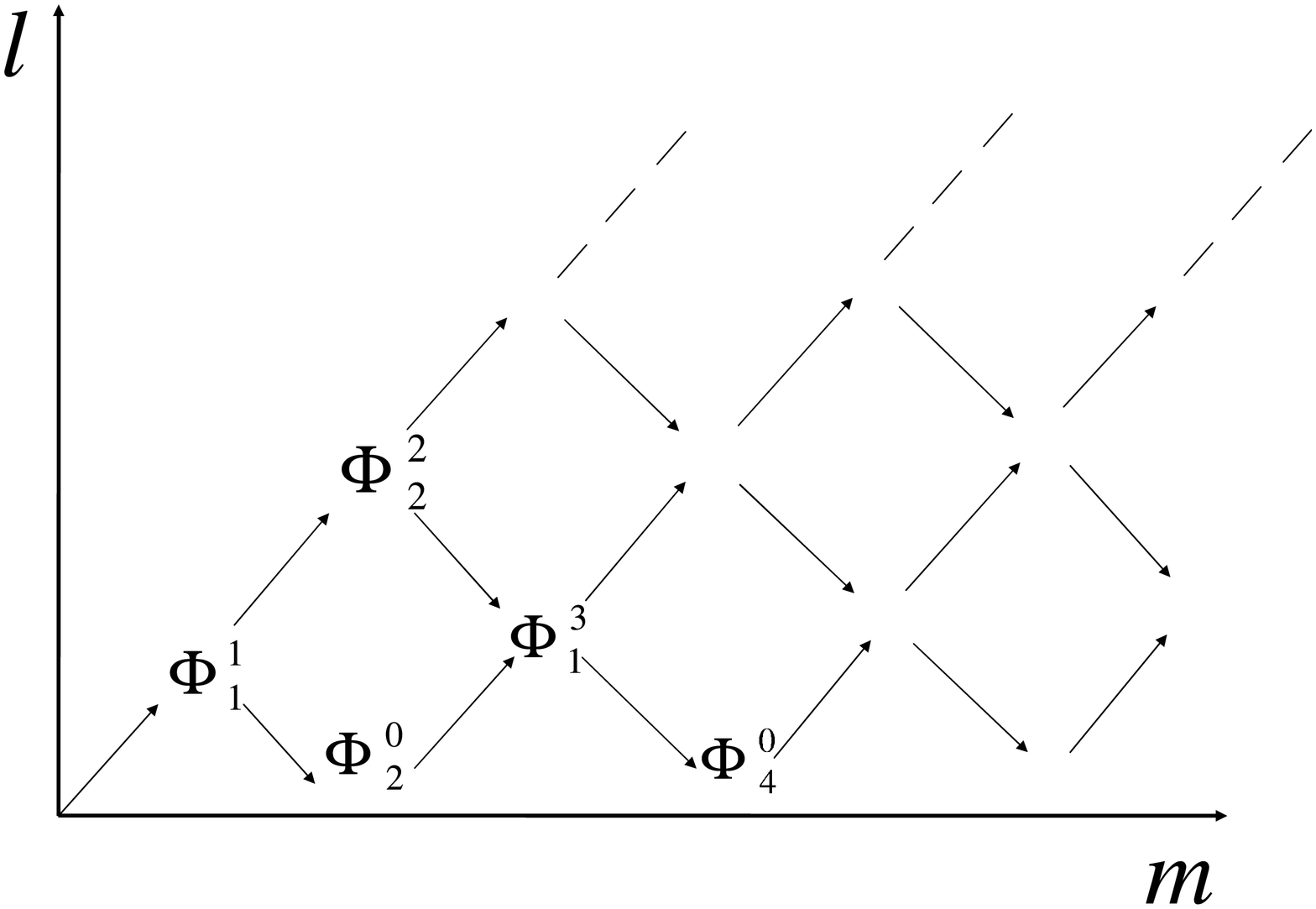}
\vspace{5mm}\caption{\label{Brattelli2} A Brattelli diagram for a Read-Rezayi state. Each node on the diagram describes the state that the quasi-particles fuse to. Each trajectory on the diagram is a ground state. }
\end{figure}

The fusion rules Eq. (\ref{eq:fusion}) introduce the non-abelian side of these quasi-particles, since as we fuse many $\sigma_1$'s together, we get a number of fusion channels that increases exponentially with the number of fused $\sigma$'s. Each of these channels corresponds to a ground state. For keeping track of this number it is useful to use again the Brattelli diagram, see Fig. (\ref{Brattelli2}). In the Brattelli diagrams that are useful to us the $x$-axis counts the number of quasi-particles (which, when divided modulo $2k$, gives the topological charge). The $y$ axis is the index $l$. A fusion with $\sigma_1$ increases or decreases $l$  by one, and $l$ is limited between $0$ and $k$, and hence the structure reflected in the figure: the diagram has $k+1$ "floors" and each steps towards the right or left takes you one floor up or down. A trajectory that starts at the origin and ends at some node of a particular number of quasi-particles corresponds to a ground state with that number of quasi-particles. Each node in the diagram corresponds to a field $\Phi_m^l$. The node at which a trajectory ends corresponds to the state to which the quasi-particles fuse in that particular ground state.

The Brattelli diagram makes the counting of the number of ground states relatively simple - it becomes the counting of trajectories on a graph. For a large number $N$ of quasi-particles, it scales like
\begin{equation}
\left[2\cos{\frac{\pi}{k+2}}\right]^N
\label{degeneracy}
\end{equation}
To propagate horizontally on the diagram, we need to fuse a state with $\sigma_1$'s. To propagate vertically, we need to fuse with the field $\Phi_0^2$. 

The information we gave here on the CFT formulation of the Read-Rezayi gives most of the background needed for generalizing the analysis of interferometry, described above in detail for the $\nu=5/2$ state, to these states. We will not carry out this generalization here, but rather refer the reader to \cite{Bonderson06b,Bonderson07,law-2007,
Ilan07,Fidkowski07}. As shown in these works, lowest order interference in the Fabry-Perot interferometer and Couomb blockade peak spacings in a quantum dot reflect the number and state of the localized quasi-particles in the cell of the interferometer, and the effective charge in a shot noise measurement in a Mach-Zehnder interferometer may identify the $Z_3$ nature of the $\nu=12/5$ state.

\section{Summary}

Much of physics is about setting rules, and then looking for ways to break them. Anyons were the breaking of the rule that the wave function of identical particles must follow fermionic or bosonic symmetry. The breaking of this rule signaled the door to a rich and fascinating part of physics, in which  ohmic contacts and semi-conductors mingle with topology and CFT's, and the immensely complicated nature of the first two converges in certain limits to the pure simplicity of the last two. Much is presently understood about that part of physics, as we tried to describe. Much more is awaiting for future research, we believe. 

%

\section{Acknowledgements}
I am indebted to Merav Dolev, Eytan Grosfeld, Roni Ilan and Gil Tayar for their help in preparing this manuscript. This work was partially funded by the US-Israel Binational Science Foundation, the Minerva Foundation and the Israel Science Foundation.

%
%
%
%
%

\end{document}